\documentclass[twocolumn]{aastex631}

\newcommand\bj{B\&J}
\newcommand\hd{HALO7D-X}
\newcommand\feh{[Fe/H]}
\newcommand\alp{[$\alpha$/Fe]}
\newcommand\lcdm{$\Lambda$CDM}
\newcommand\msun{$\rm{M}_{\sun}$}
\newcommand\Gaia{{\sl Gaia}}
\newcommand\Galaxia{{\sl Galaxia}}
\newcommand\vlos{$v_{\rm{LOS}}$}
\newcommand\vtd{$v_{\rm{3D}}$}

\usepackage{mathtools}
\usepackage{amsfonts}
\usepackage[shortlabels]{enumitem}
\usepackage{xcolor}
\usepackage{soul}

\shorttitle{The \hd\ Survey}
\shortauthors{Apfel et al.}

\received{September 1, 2024}
\revised{April 25, 2025}
\accepted{June 11, 2025}
\submitjournal{The Astrophysical Journal}

\graphicspath{{./}{figures/}}

\begin{document}

\title{Constraining the Milky Way Halo Accretion History With Simulated Stellar Halos:\\ Designing the HALO7D-X Survey}

\correspondingauthor{Kevin McKinnon}
\email{kevin.mckinnon@utoronto.ca}

\author[0009-0001-2827-1705]{Miranda Apfel}
\affiliation{Department of Astronomy \& Astrophysics, University of California, Santa Cruz, 1156 High Street, Santa Cruz, CA 95064, USA}

\author[0000-0001-7494-5910]{Kevin McKinnon}
\affiliation{David A. Dunlap Department of Astronomy \& Astrophysics, University of Toronto, 50 St. George Street, Toronto, ON M5S 3H4, Canada}
\affiliation{Canadian Institute for Theoretical Astrophysics, University of Toronto, 60 St. George Street, Toronto, ON M5S 3H8, Canada}
\affiliation{Department of Astronomy \& Astrophysics, University of California, Santa Cruz, 1156 High Street, Santa Cruz, CA 95064, USA}

\author[0000-0002-6667-7028]{Constance M. Rockosi}
\affiliation{Department of Astronomy \& Astrophysics, University of California, Santa Cruz, 1156 High Street, Santa Cruz, CA 95064, USA}

\author[0000-0001-8867-4234]{Puragra Guhathakurta}
\affiliation{Department of Astronomy \& Astrophysics, University of California, Santa Cruz, 1156 High Street, Santa Cruz, CA 95064, USA}

\author[0000-0001-6244-6727]{Kathryn V. Johnston}
\affiliation{Department of Astronomy, Columbia University, 550 West 120th Street, New York, NY, 10027, USA}

\begin{abstract}
We present the design for \hd, a survey of the stellar halo to investigate the accretion history of the Milky Way. The survey will use a combination of Hubble Space Telescope (HST) and \Gaia\ data for sky position and proper motions of faint stars ($18<G<21.5$~mag), while line-of-sight velocity, distance, \feh, and \alp\ will be measured using follow-up Keck spectroscopy. The survey will cover 30 lines of sight, made up of multiple HST archival fields and optimized for Keck DEIMOS spectroscopy. We use mock survey observations of the Bullock and Johnston stellar halo simulations to investigate the sensitivity of \hd\ to constrain the basic parameters of the accretion history of our Galaxy's stellar halo. We find that we are sensitive to the mass distribution and accretion timeline of the stellar halo progenitors, but not their orbital circularity. We find that the simulated halos fall into three different groups based on the similarities in their distributions of the observable dimensions of our survey. These groups are also distinct from each other in the mass distribution and accretion timeline of their progenitor satellites, showing that by using similarities in our observables among halos, we are able to identify similarities in their accretion histories. With \hd\ we will compare real Milky Way data with simulated halos and use this connection between observables and progenitor mass and accretion timeline to learn about the formation of our Galaxy's stellar halo.

\end{abstract}


\section{Introduction} \label{sec:intro}
The Milky Way (MW) galaxy is host to many smaller satellite dwarf galaxies, and, as these companions are accreted, their stars are tidally stripped away and become part of the MW \citep[e.g.,][]{searlezinn1978, whitefrenk1991}. Through this process, much of the stellar mass of the MW halo was built up, while the rest likely formed \textit{in-situ} in the disk and was subsequently kicked up during mergers \citep[e.g.,][]{purcell2010, sheffield2012, bonaca2017, belokurov2020}. Simulations predict that the stellar remains of the satellite accretion process should be detectable as substructure in the MW halo \citep[e.g.,][]{johnston1996,helmiwhite1999,bullock2001, abadi2006, cooper2010}. There is now ample observational evidence of these remains of past merger events \citep[e.g.,][]{ibatagilmoreirwin1995, newberg2002, belokurov2006, belokurov2007a, juric2008}. To date, nearly 100 stellar streams have been identified in the MW halo \citep[e.g.,][]{newberg2002,grillmair06a, grillmair06b, grillmair09, grillmair14, martin14, shipp2018, malhan2018, ibata2021} as well as more diffuse clouds and plumes \citep[e.g.,][]{belokurov2007b,juric2008,vivas2008,donlon2019}. For a complete list of known streams, see \textit{galstreams} \citep{mateu2023}. An ongoing merger with the dwarf spheroidal Sagittarius galaxy \citep{ibatagilmoreirwin1994}, as well as evidence for a massive merger event $\sim 10$ Gyr ago \citep[e.g.,][]{helmi2018,belokurov2018} contribute to our understanding of the stellar halo as built up of accreted stars. 

The MW is not the only galaxy in which tidal features have been found. Stellar streams and remnants of disrupted dwarfs have been found in  M31, M33, M32, NGC 205, NGC 147, and more \citep{ibata2001, carlin2016}, evidence for the importance of this process for galaxy evolution.  Although the stellar halo is a small portion of the total stellar mass of the MW \citep[e.g.,][]{deason2019, mackerethbovy2020}, it has long been of interest because of the wealth of information it contains about the history of the Galaxy.

Due to long dynamical times in the halo, accreted stars retain information about their progenitor satellites in their kinematics as well as their chemical abundances even after their spatial coherence is lost. The stars in the stellar halo have spatial distributions, 3D kinematics, and abundances that are the product of the progenitor satellite star formation histories, orbits, and the merging process that together comprise the host galaxy halo accretion history. Stellar halos with similar accretion histories will have corresponding similarities in their observable properties. Indeed, \citet{bj08} compares simulations of a galaxy artificially constructed from entirely recent accretion versus one made up entirely of ancient events and finds they are observably different. The goal of this paper is to determine how well the \hd\ survey -- an expansion of the original HALO-7D survey \citep{c19a,c19b,mckinnon23a} -- can distinguish between different accretion histories for the MW, and how this survey can be used to investigate the MW formation history and growth.

The \hd\ survey data will consist of proper motions measured by combining HST and \Gaia, and spectroscopy with Keck II/DEIMOS, across 30 lines of sight selected from the HST archive.  From these measurements, we can determine 3D positions, 3D kinematics, and chemical compositions for Milky Way stars at Galactocentric distances $>$ 15 kpc. This will give us seven dimensions of information for stars in the MW halo. In this paper, we describe the \hd\ survey design and parameters. We also present a technique to compare our \hd\ observations to simulated MW-like galaxy stellar halos to determine which of the main parameters of our Galaxy's accretion history -- such as the timeline and mass distribution of the progenitors -- we can learn from our survey.    

We now have an abundance of kinematic and chemical data for stars outside the Solar neighborhood. \Gaia\ \citep{gaia2016} has provided full-sky parallaxes and proper motion measurements, and millions of the same stars have been observed by spectroscopic surveys like H3 \citep{conroy2019}, APOGEE \citep{majewski2017}, SEGUE \citep{yanny2009, rockosi2022}, RAVE \citep{steinmetz2006}, GALAH \citep{desilva2015}, DESI \citep{DESI2016}, and LAMOST \citep{cui2012}.

Most of the local stellar halo is dominated by the remnant stellar population of a single massive accretion event $\sim~10$ Gyr ago: the ``Gaia-Sausage'' or Enceladus, GSE \citep[e.g.,][]{helmi2018,belokurov2018}. Inside 20 kpc, 50\%--70\% of the stellar halo is made up of GSE stars \citep{lancaster2019b, naidu2020}.

Outside of the local halo, the most significant ongoing merger is that of the Sagittarius dwarf spheroidal galaxy \citep{ibatagilmoreirwin1994}, which is currently being tidally disrupted as it orbits the MW. This merger has been linked to a significant amount of the substructure in the outer part of the halo, but does not have a presence in the local halo \citep[e.g.,][]{belokurov2006, koposov12, naidu2020}.

At larger distances, the halo becomes less homogeneous. Studies using Main Sequence turnoff (MSTO) stars, K giants, and blue horizontal branch (BHB) stars show that most of the distant halo is made up of substructure \citep[e.g.,][]{bell2008, starkenburg2009, xue2011, schlaufman2012, janesh2016, deason2018}. The original HALO7D survey found there are significant variations in velocity anisotropy \citep{c19b} and chemical abundances \citep{mckinnon23a} across distant lines of sight in the distant halo. This is supported by simulations that show there should be more progenitors making up a significant fraction of the halo with increasing distance \citep[e.g.,][]{bj08, fattahi2020}.  This is our motivation to observe the distant halo along many more lines of sight in the \hd\ Survey.

Many observational surveys of the MW stellar halo have either looked for remnants of specific accretion events \citep[e.g.,][]{bonaca2020, bonaca2021, cook2022, malhan2022}, or statistically quantified the amount of substructure \citep[e.g.,][]{bell2008,janesh2016,lancaster2019a}. \citet{cunningham2022} model simulated stellar halos with template chemical abundance distributions and find they can use the best-fit models to these chemistry observables to constrain properties of the progenitor satellite population and the accretion history of the simulated halos. This paper investigates our power to use 7-dimensional data with the \hd\ survey to measure the important parameters that describe our Galaxy's accretion history: when the MW distant stellar halo was accreted and the distribution of orbits and masses of the progenitors. We use simulated Milky-Way mass stellar halos with different accretion histories as templates against which to compare our \hd\ observations of position, velocities, and chemistry.  We create mock \hd\ observations of these simulated halos, and use them to build probability distributions of our observables for each halo. We compare the simulated halos to one another in order to see what parameters of their accretion histories are shared by halos that are statistically similar in our observables.

This paper is organized as follows: \S \ref{sec:halo7d} discusses the design of the \hd\ survey. In \S \ref{sec:sims} we review the \citet{bj05} simulated stellar halos, the \Galaxia\ code we use to post-processes the simulations to observables, and the process by which we create mock \hd\ observations. In \S \ref{sec:statmodels} we introduce the multi-dimensional statistical models we create from the simulated halos and the process of evaluating mock \hd\ observations against the models. In \S \ref{sec:results}, we show the results of our analysis, evaluate our survey design, and describe our sensitivity to halo accretion history parameters. We summarize our conclusions in \S \ref{sec:end}.

\begin{figure*}[ht!]
\plotone{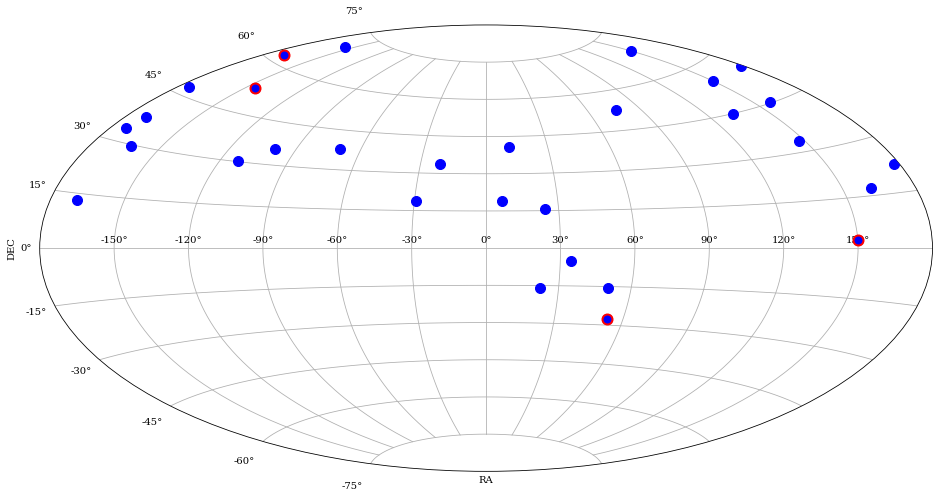}
\caption{The locations of all 30 \hd\ LOS. Fields from the original HALO7D survey are indicated with a red ring. The new survey increases the area observed. The LOS sizes are not to scale. \label{fig:LOS}}
\end{figure*}

\section{The \hd\ Survey}\label{sec:halo7d}

The goal of the \hd\ survey is to obtain seven-dimensional data for distant Milky Way halo stars along many more lines of sight (LOS) than was possible in the original HALO7D survey. The limiting observational dimensions are the proper motions (PM). The HST deep-mosaiced, multi-epoch coverage used to measure PMs of faint stars ($19 < G < 24.5$~mag) for the original HALO7D survey is limited to just a few fields. The PM measurement uncertainties near the \Gaia\ faint limit (e.g., $G\sim21$~mag) are too large to provide good kinematic constraints in an individual $<1\degr$ line of sight suited for deep spectroscopy with DEIMOS on Keck. However, there have been promising developments in combining HST and \Gaia\ data to measure more precise PMs \citep[e.g.,][]{massari17, delpino2022, warfield2023}. These HST+\Gaia\ PMs have been used to study many MW globular clusters and satellite galaxies \citep[e.g.,][]{massari18, bennet2023}. \citet{mckinnon23b} extended this technique to sparse fields (i.e., fewer than 10 \Gaia\ stars per HST image) and showed that combining the two observatories allows us to measure PMs for Milky Way field stars with $G \gtrsim 18$~mag more precisely than \Gaia\ alone. Depending on the time baseline between the HST and \Gaia\ observations in sparse fields, \citet{mckinnon23b} can measure PMs that are $\sim5-20$ times more precise than \Gaia\ alone for $G=21$~mag and recover PMs for fainter stars where \Gaia\ is more incomplete (i.e., $21<G<21.5$~mag). Using this method, \hd\ will be able to measure the PMs of stars at larger distances and on more LOS in the HST archive as compared to the original HALO7D survey in order to investigate the formation of the MW halo. 

With this technique for measuring PMs, the possible observation fields for spectroscopic follow-up expand to anywhere that HST has imaged at least 4~yr before the end of data collection for \Gaia\ DR2 on 2016 May 23, our minimum baseline for the PM measurements used in our design of \hd.  Starting with the entire HST exposure catalog before that date, we selected fields that (a) had $-30\degr<\delta<+75\degr$ so as to be accessible with the Keck telescope for the spectroscopic component of the survey, (b) were taken with either WFC3/UVIS or ACS/WFC, detectors with large field of view, small pixel size, and high efficiency, (c) used one of the filters F555W, F606W, F775W, F814W, or F850LP, which have the best astrometric calibrations available as well as optimal S/N, (d) had exposure times between 50 and 500~s, to have high S/N while not saturating \Gaia\ stars, (e) had Galactic latitude $|b_{\rm Gal}|>5\degr$ to mitigate against crowding, (f) had ecliptic latitude $|b_{\rm Ecl}|>5\degr$ so as to avoid exposures that tracked Solar System bodies, and (g) had $A_V <2$ to remove regions with high dust extinction. Many exposures had duplicate or near duplicate pointings, so we condensed those that overlapped by at least 1\arcmin\ into a single {\it field\/} location, leaving us with $\sim1500$ unique fields. 

\begin{figure}[h]
\plotone{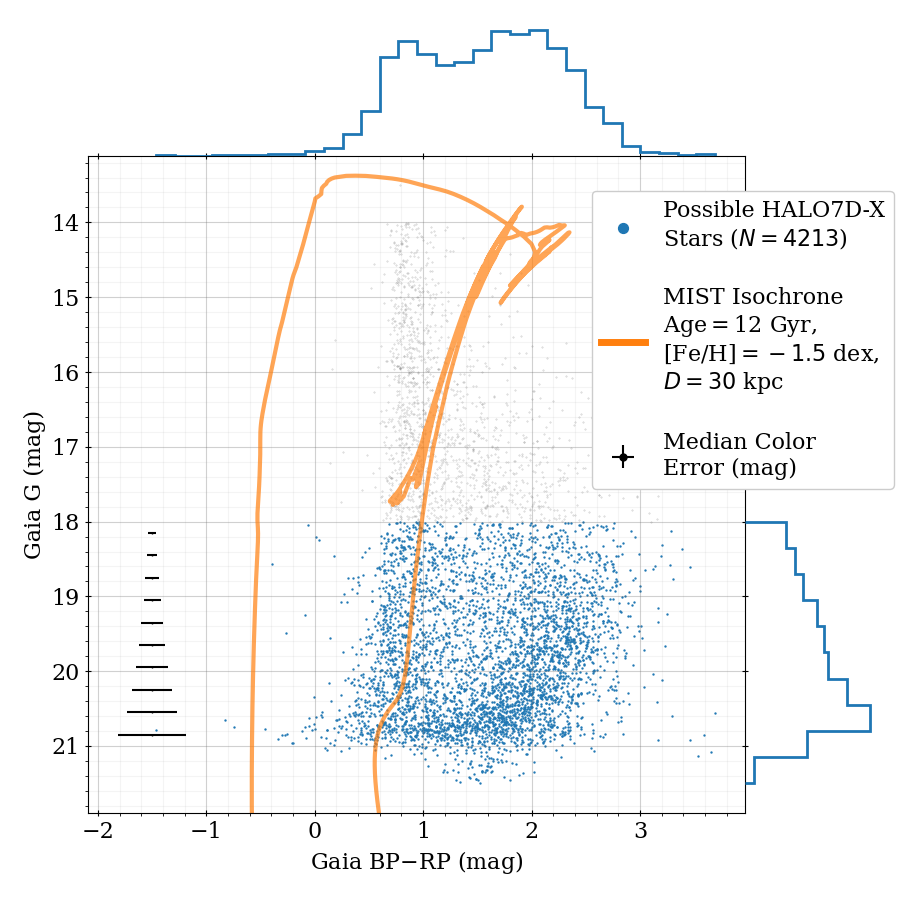}
\caption{Color magnitude diagram of the $\sim 4200$ Gaia stars in the COSMOS field that could potentially be included in a future \hd\ survey (blue points). Median color uncertainties for different magnitude bins are shown as black error bars at BP$-$RP$=-1.5$~mag, and histograms of the magnitude and color are displayed along the top and right edges. A typical MW stellar halo MIST isochrone \citep{Dotter_2016,Choi_2016,Paxton_2011,Paxton_2013,Paxton_2015} is shown in orange to guide the eye. Grey points are stars that also fall in the COSMOS region, but are excluded from potential \hd\ samples.
\label{fig:COSMOS_CMD}}
\end{figure}

We estimate each field ($160\arcsec \times 160\arcsec$ for WFC3/UVIS, $202\arcsec \times 202\arcsec$ for ACS/WFC) to contain an average of 5 confirmed MW halo stars within our magnitude limits, which is too few for our analysis. Therefore, we identified all groupings of at least 3 fields where each field was separated from its closest neighbor by no more than 16\arcmin\ (the length of a DEIMOS mask), to increase the efficiency of the spectroscopic observations. Each of these groups of fields has small enough inter-field separations that they can be considered together as a single LOS. We visually inspected these groupings and removed those centered on bright, extended objects such as the Magellanic Clouds and large other nearby galaxies. This resulted in a final \hd\ survey design of 30 LOS, covering 371 unique fields, with an expectation of at least 15 observable halo stars for each LOS. However, many of the LOS will have significantly more halo stars than this, such as the original HALO7D fields. The LOS locations are shown in Figure~\ref{fig:LOS}, and an example CMD of Gaia stars in the COSMOS field  \citep[(RA, Dec) $\approx (150.1^\circ,2.2^\circ)$;][]{Nayyeri_2017,Muzzin_2013} is shown in Figure~\ref{fig:COSMOS_CMD}.

This survey was conceived as an extension of the original HALO7D survey, as described in \citet{c19a,c19b,mckinnon23a}. Although the limiting magnitude of \hd, set by the faint \Gaia\ limit, is not as deep as the original survey, we compensate by increasing the number of LOS by roughly an order of magnitude. Included in our 30 LOS are expansions of the four original HALO7D fields, expanded because we can now include single-epoch HST pointings beyond the smaller sub-regions of those fields with multi-epoch HST data.

For the \hd\ extension, our observable parameters are on-sky position and PMs using HST+\Gaia\ as in \citet{mckinnon23b}, LOS velocity measured as described in \citet{c19a}, and distance, \feh, and \alp\ from Keck spectroscopy \citep[e.g.][]{Cargile_2020,mckinnon23a}. 

\section{Simulated Halos}\label{sec:sims}

\begin{figure*}[ht!]
\plotone{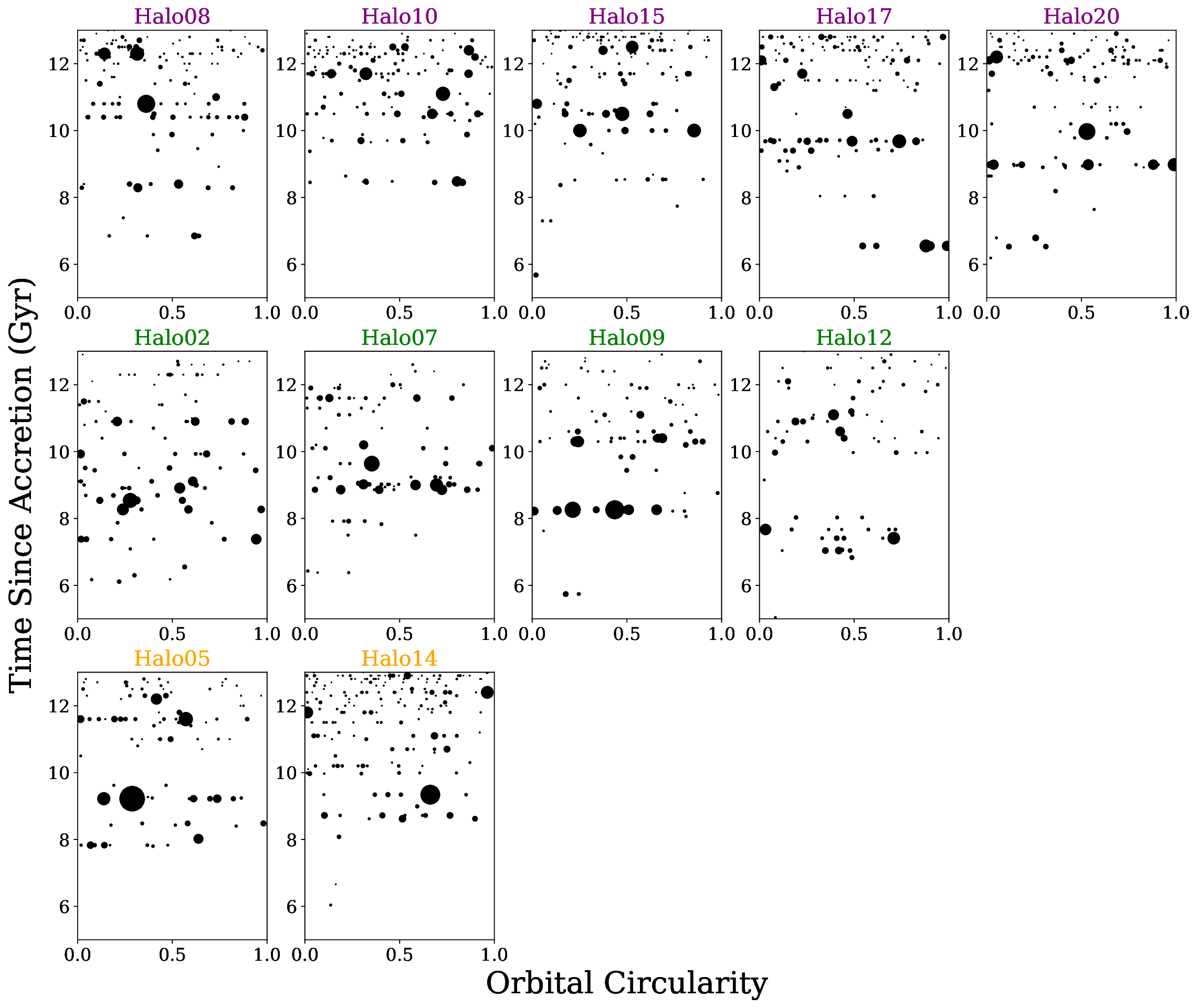}
\caption{The individual accretion histories of the 11 different \bj\ halos. Each symbol represents an individual accreted satellite, with the size of the point linearly scaled by the dark matter mass of the satellite. The x-axis of each plot shows the orbital circularity of the satellites, with 0 indicating a perfectly radial orbit and 1 a perfectly circular one. The y-axis is the time elapsed since the satellite was accreted by the host galaxy, with the most ancient events at the top. Satellites in the simulation that are still intact (i.e., bound by their self-gravity) are not included. The halos are divided into three rows and the halo number title of each panel is color coded by the the distinct groups identified and described in \S \ref{sec:simdist} and Figure~\ref{fig:2dhist}.\label{fig:history}}
\end{figure*}

We now investigate how much we can learn from our \hd\ survey about the timeline, progenitor mass, and satellite orbital parameters of our Galaxy's accretion history. We do this by using mock observations of simulated stellar halos with a range of different but known accretion histories.

We use the Bullock and Johnston suite of galaxy halo simulations \citep{bj05, bj08, font06a, font06b, robertson05}, hereafter \bj\ halos. The suite consists of 11 simulated halos with different accretion histories, all built within the context of a \lcdm\ universe. They are a hybrid of semi-analytic and \textit{N}-body simulations.

The simulations use a three-component potential for the host galaxy made up of the disk, bulge, and spherically symmetric dark matter halo. The evolution of the dark matter component of each satellite is followed self-consistently with an \textit{N}-body code, while the host galaxy components are represented by smoothly evolving analytic functions.

In order to produce a realistic stellar halo, the simulations explicitly distinguish between the evolution of dark and luminous matter in accreted satellites. The halo stars modeled are only those originating from accreted dwarf galaxies, by ``painting'' the dark matter particles with appropriate stellar masses. Therefore, the simulated halos do not include contributions from globular clusters, nor any stars that were formed \textit{in-situ} or kicked up from the disk. The resulting stellar halos have similar density profiles and total masses to the MW. The satellites have comparable structural parameters to the classical dwarf spheroidal satellites of the MW (i.e. those that were known at the time of publication of the \bj\ models), and there are similar numbers of high-surface-brightness surviving satellites.

The \bj\ halo chemical enrichment was modeled using a leaky-accreting-box model \citep{robertson05}, with the parameters adjusted to reproduce both the low $\alpha$-element ratios of MW satellites and the Local Group dwarf mass-metallicity relationship. The resulting stellar halos have abundance patterns similar to stars of the MW halo \citep{font06a}.

The simulations become more accurate for the more recent accretion events, and therefore the further out into the halo ($\geq20$ kpc), since the inner halo is built up from early accretion. None of the halos have had a merger with a satellite $\geq10\%$ of the parent mass in the last 7 Gyr, which is appropriate as we have no evidence for a major merger much more recent than that in the MW. The progenitor satellite of the most dominant merger found to date, the GSE, has a mass ratio with the MW of $\sim 4:1$, and was accreted approximately 8--10 Gyr ago \citep{helmi2018}.

We use the software \Galaxia\ \citep{galaxia}, which is distributed with the \bj\ halos, to make mock observations. \Galaxia\ smooths and interpolates over the \bj\ star particles, while correctly populating phase-space, to determine the properties of individual stars. \Galaxia\ provides positions, velocities, chemistry, and apparent magnitude in multiple photometric passbands for each of the interpolated stars it creates, and tracks the progenitor satellite for each star. For our analysis, \Galaxia\ did not observe satellites that are still intact. \hd\ will only observe halo fields without any known dwarf satellite galaxies or star clusters, as we are interested in properties of stars that have already been accreted into the halo.

\subsection{Accretion History of Simulated Halos}\label{sec:char}
Figure~\ref{fig:history} shows the accretion histories of the 11 \bj\ halos: the time of accretion, orbital circularity, and mass of each progenitor satellite galaxy. We expect each of the accretion history parameters included in the figure to have a significant effect on the observable properties of the stellar halos. Halos that have many recent accretion events have a higher number of substructure features than halos with few or no recent accretion events and a higher fraction of the total halo stars in these features. Even after stars from a progenitor disperse over time, losing their positional cohesion, we can still identify these features as dynamically cold peaks in the velocity distribution, and as concentrations in the distribution of chemical abundances. In the simulations, it is assumed that star formation stops soon after a satellite galaxy is accreted. This means that the length of the star formation history of a satellite is determined by the epoch of its accretion into the parent halo. Ancient accretion events are associated with satellites that have short star formation histories, and their chemical enrichment is dominated by Type II supernovae, leading to high \alp\ in their stellar population. Recent events are associated with satellites that have longer star formation histories with time for contributions from both Type II and Type Ia supernovae, leading to lower values of \alp. The orbital circularity of a satellite determines the morphology and physical scale of the resultant substructure feature as it is tidally disrupted. Circular orbits often lead to narrow stream-like features, while radial orbits tend to result in larger clouds, plumes, and shells. The mass of each progenitor affects both the physical scale of the resulting feature and \feh, since the simulated satellites follow the present-day mass-metallicity relation for dwarf galaxies, with lower masses corresponding to lower values of \feh.

\subsection{Mock \hd\ Surveys}\label{sec:simdata}

To quantify how well our \hd\ survey can distinguish the imprint of a halo's accretion history on our observables, and how that sensitivity changes with the parameters of our survey design, we create a set of mock \hd\ surveys in each of the \bj\ halos. We use these to construct multi-dimensional distribution functions that capture the relationship between each of our observables in each halo. These will become the probability mass functions described in \S \ref{sec:models}.

To capture the full range of our observables in each of the halos,  we take sets of \Galaxia\ mock observations of our 30 LOS with the Sun at 360 different locations in the plane of the disk. For each observation, we move the Sun in a circle around the simulated halo center, while keeping the distance from the center and height above the disk mid-plane constant.

For each mock \hd\ survey in \Galaxia\ at a single Solar location, we select all stars in our expected spectroscopic magnitude range of $18 < V < 21.5$ in all of our 30 LOS defined in \S \ref{sec:halo7d}.  We model each LOS with \Galaxia\ as a circle with area 1 square degree centered on the LOS coordinates. Our actual LOS are smaller in area and are made up of often non-contiguous sets of individual HST fields, which can each be separated by up to 16\arcmin. We chose 1 square degree as it is large enough to contain any of the bounded areas of our fields, while still being smaller than the angular size of substructure features in the \bj\ halos at the distance of our \hd\ data.  

In \S \ref{sec:num}, we vary the number of lines of sight and number of observed stars per field in our \hd\ survey to investigate how that changes our ability to distinguish between different halo accretion histories. For that analysis, we draw subsets of the LOS and subsets of stars in each LOS from the mock observations described here. 

With \hd, our basic observational data are sky position, the two components of proper motion, LOS velocity (\vlos), distance, \feh, and \alp. Using our distance estimates, we can convert the PMs into transverse velocities, although in the real observations this increases the uncertainty in our kinematic measurements (see \S \ref{sec:obserrors}). The three velocity components are converted into the Galactic Standard of Rest (GSR) frame by removing the component of the Sun’s peculiar motion along each direction. The velocity components are then combined into a single 3D velocity, \vtd. 

We use combinations of these parameters to build our statistical models. The baseline set of observables we use to investigate the sensitivity of our \hd\ survey to halo accretion history is \vtd, distance modulus $\mu$, \feh\ and \alp.  We also investigate model distributions using subsets of these observables, and \vlos\ instead of \vtd, in \S \ref{sec:obserrors}.

\section{Statistical Recovery of Halo Accretion History}\label{sec:statmodels}

Our goal is to understand what we can learn about the parameters of our Galaxy's accretion history from \hd. We create mock \hd\ surveys with different design configurations, and, for each mock survey, we compare the observables to the distributions of those same observables from the full set of simulated halos. We then ask the following questions: 
\begin{enumerate}
    \item For a mock survey and its observables, can we determine the simulated halo from which those data were drawn?
    \item How does survey design (number of LOS, number of stars per LOS, choice of observables) impact the ability to correctly identify the origin halo? 
    \item Can quantitative comparison of the mock survey stellar observables to the suite of simulated halos reveal information about the accretion history of the origin halo?
\end{enumerate}

We create probability mass functions by gridding the observables in our mock surveys (\S \ref{sec:models}) and then use those distributions to answer the questions listed above (\S \ref{sec:sampeval}).

\subsection{Evaluating Probabilities}\label{sec:models}
For each simulated halo, we build a multidimensional probability distribution on the observables using different combinations of 3D velocity, LOS velocity, distance modulus, \feh\ and \alp.  (\S \ref{sssec:pdfs}). Once we have these probability distributions, we can evaluate the likelihood of a star's observables originating from each halo (\S \ref{sssec:eval_sim_data}). This section of the paper will make extensive use of the variable definitions shown in Table~\ref{tab:statistics_definitions}. 

\begin{table*}[t]
\caption{Definitions for Statistical Model
\label{tab:statistics_definitions}}
\centering
\begin{tabular}{c|c|c}
\hline \hline
Variable     & Description      & Properties                   \\ \hline
$n_{\rm LOS}$ & Number of data LOS & $5 \leq n_{\rm LOS} \leq 30$  \\
$i$ & Data LOS index & $i \in \{1, 2, \dots, n_{\rm LOS} \}$ \\
$n_*$ & Number of stars per LOS & $5 \leq n_* \leq 100$ \\
$j$ & Star index & $j \in \{1, 2, \dots, n_* \}$\\
$n_T$ & Number of simulated LOS & $n_T \geq n_{\rm LOS},\, n_T = 30$ \\
$k$&  Simulated LOS index & $k \in \{1, 2, \dots, n_T \}$ \\
$n_{\rm cell}$ & Number of probability cells in grid & $n_{\rm cell} = 3360$ \\
$l$  &  Probability cell index & $l \in \{1, 2, \dots, n_{\rm cell} \}$\\
$n_{\rm halo}$ & Number of simulated halos & $n_{\rm halo} = 11$ \\
$h$  &  Simulated halo index & $h \in \{1, 2, \dots, n_{\rm halo} \}$\\
$n_{\rm ang}$ & Number of angles the Sun was placed in & $n_{\rm ang} = 360$ \\
$s$  &  Simulated Solar angle index & $s \in \{1, 2, \dots, n_{\rm ang} \}$\\
$\vec \theta_{i,j}$ & Vector of stellar observables for star $j$ in data LOS $i$ & e.g., $\vec \theta_{i,j}^T = \left(v_{\mathrm{3D} ,i,j},~ \mu_{i,j},~ \mathrm{\feh}_{i,j},~ [\alpha/\mathrm{Fe}]_{i,j} \right) $ \tablenotemark{a}\\
$\pmb d_{i}$ & Matrix of stellar observables for all stars in data LOS $i$ & $\pmb d_{i} = \left(\vec \theta_{i,1},~ \vec \theta_{i,2},~ \dots,~ \vec \theta_{i,n_*} \right)$ \\
$\pmb D$ & Matrix of stellar observables for all stars in all data LOS & $\pmb D = \left(\pmb d_{1},~ \pmb d_{2},~ \dots,~ \pmb d_{n_{\rm LOS}} \right)$ \\
$x_{h,k,s,l}$ & Probability in cell $l$ for simulated LOS $k$  & $\sum_{l=1}^{n_{\rm cell}}x_{h,k,s,l} = 1$\\
 & when the Sun is at angle $s$ in halo $h$ & \\
$\vec x_{h,k,s}$ & Probability vector for Solar angle $s$, simulated LOS $k$, halo $h$ & $\vec x_{h,k,s}^T = \left(x_{h,k,s,1},~ x_{h,k,s,2},~ \dots,~ x_{h,k,s,n_{\rm cell}}  \right)$ \tablenotemark{a}\\
$\pmb X_{h,k}$ & Probability matrix for simulated LOS $k$, halo $h$ & $\pmb X_{h,k}^T = \left( \vec x_{h,k,1},~ \vec x_{h,k,2},~ \dots,~ \vec x_{h,k,n_{\rm ang}} \right)$ \tablenotemark{a} \\
$\pmb \chi_{h}$ & Probability matrix for halo $h$ & $\pmb \chi_{h}^T = \left(\pmb X_{h,1},~ \pmb X_{h,2},~ \dots,~ \pmb X_{h,n_{T}}  \right)$ \tablenotemark{a}\\
$f_h$ & Probability that the data come from halo $h$ & $\sum_{h=1}^{n_{\rm halo}} f_h =  1$\\
$\vec f$ & Halo fraction vector & $\vec f^T = \left( f_1,~ f_2,~ \dots,~ f_{n_{\rm halo}} \right)$ \tablenotemark{a}\\
\hline
\end{tabular}
\tablenotetext{a}{The superscript $T$ denotes the transpose of a vector or matrix.}
\end{table*}

\subsubsection{Probability Distributions} \label{sssec:pdfs}

\begin{figure}[t]
\begin{center}
\begin{minipage}[c]{\linewidth}
\includegraphics[width=\linewidth]{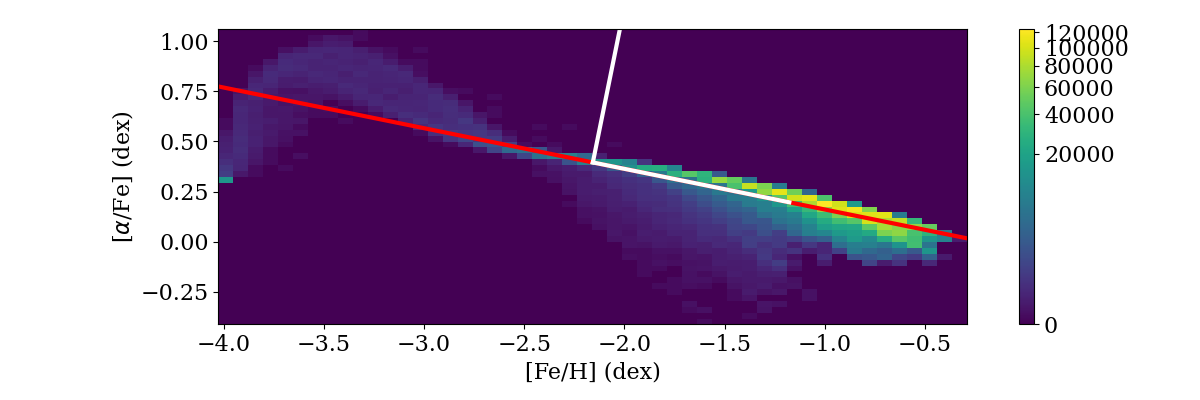}
\end{minipage}\\
\begin{minipage}[c]{\linewidth}
\includegraphics[width=\linewidth]{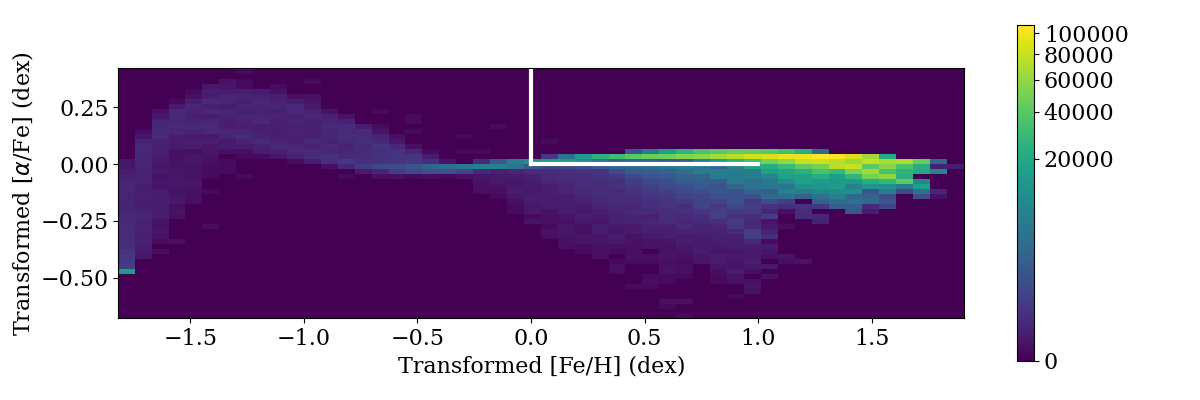}
\end{minipage}
\caption{\textbf{Top:} 2D histogram of the simulated chemical abundances from the \bj\ halos. A best-fit line is shown in red, which defines a set of perpendicular vectors (white). \textbf{Bottom:} The transformed chemical abundances defined by the (white) perpendicular vectors in the top panel.}
\label{fig:simulated_chemistry_transformation}
\end{center}
\end{figure}

Because \feh\ and \alp\ are highly correlated, we define a new, transformed abundance plane with axes approximately parallel and perpendicular to the [Fe/H]-[$\alpha$/Fe] relationship. We do this by fitting a line to the maximum density in the simulated abundance plane from all the halos, which produces a set of perpendicular vectors that define the new, transformed abundance axes shown in Figure~\ref{fig:simulated_chemistry_transformation}. This transformation makes it so that the grid in observables we define in the next paragraph will take steps along the principal axes of the \feh\ and \alp\ abundances.

We next divide the parameter space of the observables into a regularly spaced grid. The same grid is used to build the likelihood distributions for each simulated halo, and is defined in Table~\ref{tab:observables_grid}. The widths are chosen such that there are 10 cells in each of the parameter dimensions. This minimizes the impact of Poisson noise in the counts in each cell.  There are 10000 probability cells when the observables of interest are velocity (either 3D or LOS), distance modulus, transformed \feh, and transformed \alp. We investigate the impact of our choice of cell size relative to the observational errors in \S \ref{sec:obserrors}.

\begin{table*}[t]
\caption{Observables Grid Definition \label{tab:observables_grid}}
\centering
\begin{tabular}{c|c|c|c|c|c}
\hline \hline
Observable & Units     & Grid Cell Width      & Range & Number of Steps & Smoothing Scale \\ \hline
\vtd \tablenotemark{a} & $\mathrm{km~s^{-1}}$ & $120$ & $(0,1200)$ & 10 & $60$ \\
\vlos & $\mathrm{km~s^{-1}}$ & $200$ & $(-1000,1000)$ & 10 & $60$ \\
Distance modulus $\mu$& mag & $2$ & $(4,24)$ & 10 & $0.75$ \\
Transformed\tablenotemark{b} \feh & dex & $0.2$ & $(-0.3,1.7)$ & 10 & $0.05$ \\
Transformed\tablenotemark{b} \alp & dex & $0.09$ & $(-0.5,0.4)$ & 10 & $0.025$ \\
\hline
\end{tabular}
\tablenotetext{a}{Three dimensional space velocity, computed from the proper motions, LOS velocity and distance.  Note \vtd is used in our baseline analysis. We discuss results using \vlos instead in \S \ref{sec:obserrors}}\tablenotetext{b}{The ``Transformed'' abundances refer to the rotated abundances shown in the bottom panel of Figure~\ref{fig:simulated_chemistry_transformation}.}
\end{table*}

To avoid cells with probabilities of 0 when distributing stars in the grid cells to construct the likelihood distributions,  each simulated star is modeled as the center of a multidimensional distribution in observables. Specifically, we consider the set of observables of each simulated star to be the mean of a multivariate normal distribution with a covariance matrix defined by the smoothing scales shown in the rightmost column of Table~\ref{tab:observables_grid}. The smoothing widths have been chosen to be approximately half of the 68\% width of the distribution of each observable over all simulated halos. Instead of each star falling into exactly one grid cell, some fraction of the star's total weight is assigned to all nearby grid cells. This process can be thought of as having a similar effect as uncertainties on the individual observables of each simulated star. We make this explicit comparison in \S \ref{sec:obserrors}.

To estimate the probability $x_{h,k,s,l}$ of grid cell $l$, for a given halo $h$, for a given simulated LOS $k$, for a given angle of the Sun in the Galactic plane around the Galactic center $s$, we sum the weights of the simulated stars that fall in that cell and divide by the total weight of simulated stars in all the grid cells for that configuration $(h,k,s)$. That is, the probabilities in the grid cells for a particular choice of $(h,k,s)$ sum to 1: $$\sum_{l=1}^{n_{\rm cell}}x_{h,k,s,l} = 1.$$

\subsubsection{Evaluating Probability of Data} \label{sssec:eval_sim_data}

We consider the matrix of the chosen set of observables $\theta_{i,j}$, real or mock, from a collection of stars, $j=1,n_*$, to be the data we are analysing for each data LOS $i$:
$$\pmb d_i = \left(\vec \theta_{i,1},~ \vec \theta_{i,2},~ \dots,~ \vec \theta_{i,n_*} \right).$$ For a complete survey, the matrix of all observables is $$\pmb D = \left(\pmb d_1, \pmb d_2, \dots, \pmb d_{n_{\rm LOS}} \right).$$
In our analysis, we keep track of the data LOS indices and the simulated LOS indices separately because we do not know a priori, for example, whether data LOS 1 should best be compared with simulated LOS 1 or one of the other simulated LOS. 

Using our probability distributions for the observables defined in \S\ref{sssec:pdfs}, we can measure the likelihood that an observation along one LOS comes from a particular halo $h$, Solar angle $s$, and simulated LOS $k$. To evaluate the likelihood of data with these distributions, we need only consider the number of stars with observables in each of the grid cells. That is, the matrix of observables along data LOS $i$, $\pmb d_i$, is equivalent to the vector of counts in each cell
$$\vec n_i = \left(n_{i,1},~ n_{i,2},~ \dots,~ n_{i,n_{cells}} \right),$$
where $\sum_{l=1}^{n_{\rm cell}} n_{i,l} = n_*$.  For the complete survey, these count vectors along each data LOS define the matrix of all count vectors $$\pmb N = \left(\vec n_1, \vec n_2, \dots, \vec n_{n_{\rm LOS}} \right).$$

With these definitions in hand, we see that the likelihood of the stars in data LOS $i$ originating from simulated LOS $k$ of halo $h$ with the Sun at Solar angle $s$ is a multinomial ($\mathcal{MN}$) distribution using the measured probabilities in each cell:
\begin{equation} 
    \begin{split}
        p\left(\pmb d_i | h,k,s\right) = p\left(\vec n_i | h,k,s \right) = \mathcal{MN}\left( \vec n_i | \vec x_{h,k,s}\right).
    \end{split}
\end{equation}
Next, we define the total likelihood of the data from all the $n_{\rm LOS}$ in a survey coming from a particular halo using 
\begin{equation} \label{eq:likelihood}
    \begin{split}
        p\left(\pmb D | h\right) &= p\left(\pmb N | h\right)\\
        &\propto \sum_{s=1}^{n_{\rm ang}} p(s | h) \cdot p(\pmb N | h,s)\\
        &\propto \sum_{s=1}^{n_{\rm ang}} p(s | h) \cdot \prod_{i=1}^{n_{\rm LOS}} p(\vec n_i | h, s)\\
        &\propto \sum_{s=1}^{n_{\rm ang}}  p(s | h) \cdot \prod_{i=1}^{n_{\rm LOS}} \sum_{k=1}^{n_T} p(k | h,s) \cdot p(\vec n_i | h, k, s) \\
    \end{split}
\end{equation}
where $p(s | h)$ and $p(k | h,s)$ are prior distributions that weigh the different Solar placement angles and simulated LOS, respectively.  We define these priors in \S ~\ref{sssec:priors}. The probability statements in Equation~\ref{eq:likelihood} for the likelihood marginalizes over the different possible Solar angles and possible simulated LOS when comparing the observed data to the simulations. 

Finally, the halo fraction vector, $\vec f$,  which defines the probabilities that the observed data originate from each of the simulated halos, has a posterior distribution that is given by:
\begin{equation} \label{eq:posterior}
    \begin{split}
        p\left(\vec f | \pmb D\right) &= p\left(\vec f | \pmb N\right)\\
        &\propto p\left(\vec f\right) \cdot \sum_{h=1}^{n_{\rm halo}} f_h \cdot p(\pmb N | h)
    \end{split}
\end{equation}
where $p\left(\vec f\right)$ is another prior distribution we define in the following subsection. This $\vec f$ is the main component that we are interested in measuring.

Using this formalism, we end up with a distribution on the probability that a particular set of data -- mock or real, with any number of stars along each LOS, and considering up to 30 data LOS -- originate from each simulated halo. As a simplified example which considers only two simulated halos, the posterior halo fraction might have a $\vec f~^T = \left(0.9,0.1\right)$, implying that the data look 90\% like the observables in Halo 1 and 10\% like the observables in Halo 2.

\subsubsection{Setting the Priors} \label{sssec:priors}

To finish defining the posterior probability in Equation~\ref{eq:posterior}, we need to decide on priors for the halo fractions, Solar angles, and the possible different LOS that data can lie along. We choose $p(s| h) = 1/n_{\rm ang}$ and $p(k| h, s) = 1/n_{T}$, which causes the sums over $k$ and $s$ to become averages of the likelihood over all the possible simulated LOS and Solar angles. With 360 different Solar placement angles for each simulated halo and 30 different simulated LOS for each Solar angle, we believe that this model choice captures a sense of the variation that a given halo produces in observables for different LOS and locations of the Sun. For the prior on the halo probability vector, $\vec f$, we choose to use a Dirichlet distribution with all the concentration parameters, $c$, set to 1, which means that each simulated halo has equal prior probability (i.e., $f_h = 1/n_{\rm halo}$) and that all possible combinations of $f_h$ fractions are equally likely. These priors are listed in Table~\ref{tab:priors}.

\begin{table}[t]
\caption{Prior Distributions \label{tab:priors}}
\centering
\begin{tabular}{c|c}
\hline \hline
Distribution & Form\\ \hline
$p\left(\vec f\right)$  & $\mathrm{Dirichlet}\left(\vec f~ |~ \vec c \right)$,\\
 & where $\vec{c}~^T  = \left(c_1,~ c_2,~ \dots,~ c_{n_{\rm halo}} \right)$, $c_h = 1$\\
$p\left(s | h \right)$  & $1/n_{\rm ang}$\\
$p\left(k | h,s \right)$  & $1/n_{T}$\\
\hline
\end{tabular}
\end{table}

\subsection{Measuring Posterior Halo Probabilities}\label{sec:sampeval}
With our posterior distribution defined, we are able to measure the posterior probability of a mock \hd\ survey originating from each of the simulated halos. We explore a range of survey designs by varying the number of LOS and the average number of stars along each LOS.  We draw our mock \hd\ survey data from the simulated halos. To do this, we choose a mock survey configuration of $(n_{\rm LOS}, n_*)$ and any one of the halos and then draw a mock survey from its probability distribution as constructed in \S \ref{sssec:pdfs}. For example, $(n_{\rm LOS}, n_*) = (20,15)$ means we choose one simulated halo -- say, Halo02 -- then choose one of the Solar placements. We then randomly choose 20 of the 30 simulated LOS at that Solar placement of Halo02 to draw 15 sets of observables for each of the 20 LOS from the grid described in \S \ref{sssec:pdfs}. 

We use a Metropolis-Hastings (MH) MCMC approach to sample posterior values of the halo fraction vector, $\vec f$, given the mock survey data. In particular, we initialize 500 MCMC walkers with halo fraction vectors near the prior distribution mean (i.e., all $f_h = 1/n_{\rm halo}$). At every iteration in the MCMC algorithm, we draw a new possible $\vec f$ from a Dirichlet proposal distribution and then evaluate the posterior probability of those draws using Equation~\ref{eq:posterior}. We decide whether to accept those proposed  $\vec f$ draws by comparing the posterior probability of the current $\vec f$ to the proposed one using the MH algorithm, accounting for the proposal probabilities. We repeat this process for 500 iterations, though the MCMC walkers usually converge on the posterior distribution after $\sim 100$ steps. We throw out the walker positions from the first 250 iterations, and then record the median and 68\% confidence region of the remaining samples. The resulting measurements are an estimate of the posterior probability that the current data originated from each halo and the uncertainty on that measurement\footnote{While every posterior sample of $\vec f$ will sum to 1 by definition, the sum of the posterior median $f_h$ values do not necessarily sum to 1.}. 

\begin{figure*}[ht!]
\plotone{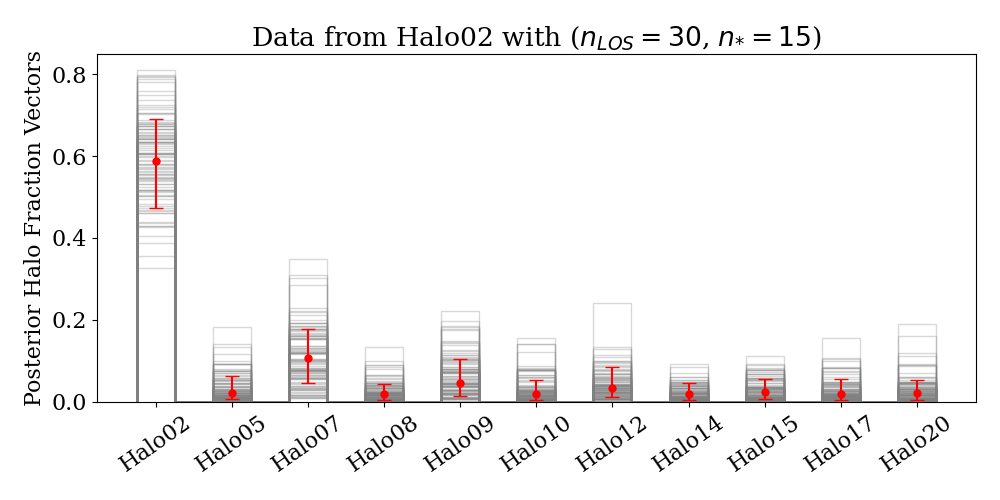}
\caption{Posterior halo fraction vectors for one data draw that originates from Halo02 and has 30 data LOS and 15 stars per LOS. The observables in this case are \vtd, $\mu$, transformed \feh, and transformed \alp. The faint grey histograms show different realizations of the posterior fraction vectors, and the red solid dot and errorbars show the posterior median and $\pm 68\%$ interval of those probabilities for each halo. The posterior probability that the data originated from the correct halo (Halo02) is significantly higher than the other halos. Halo07 has the next highest posterior fraction, implying that Halo07 has observable distributions more similar to Halo02 than the other halos. \label{fig:single_halo_prob_outputs}}
\plotone{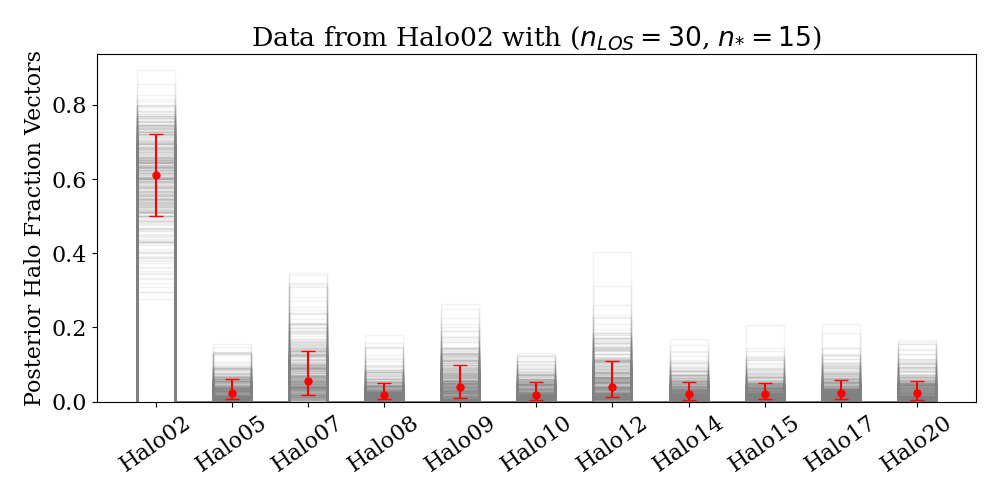}
\caption{Same as Figure~\ref{fig:single_halo_prob_outputs}, except using the results from 360 mock survey draws from Halo02 instead of just one. When we have averaged over the 360 data draws thereby minimizing the impact that any one data draw can have on the measured posterior halo fractions, we see that data from Halo02 almost always finds the highest fraction in Halo02, showing that this survey configuration is able to correctly identify when observables come from this halo. We also notice that Halo07, Halo09, and Halo12 have the next highest halo fractions, implying that data from Halo02 on average looks more similar to data from these halos than the others. \label{fig:sampeval}}
\end{figure*}

An example of output posterior halo fraction vectors for one realization of one mock survey drawn from one Solar angle of Halo02 using 30 data LOS and 15 stars per LOS is shown in Figure~\ref{fig:single_halo_prob_outputs}. We see that, for this mock survey draw, we almost always find the highest fraction for the correct origin halo. We also see that of the other halos, some (like Halo07) have higher fractions than others, implying that this drawn mock survey can potentially be described by the observable distributions of these other simulated halos. 

To marginalize over the effect that a single realization of data can have on the posterior halo fractions, we repeat the process for each survey configuration and each halo 360 times, one realization coming from each of the Solar angles used to sample the simulated halos with \Galaxia. Figure~\ref{fig:sampeval} shows this marginalized result for the same mock survey configuration illustrated in Figure~\ref{fig:single_halo_prob_outputs}: mock observations drawn from Halo02 using a configuration of 30 LOS and 15 stars per LOS. The main difference between the single realization and marginalized result is that the Halo07 peak has decreased. This suggests that the particular mock survey realization analyzed to create Figure~\ref{fig:single_halo_prob_outputs} has some pattern in its observables that are slightly more similar to Halo07 than the average set of data in this configuration from Halo02. 

We note that the maximum of the posterior halo fractions in both sets of figures is less than one. The explanation for this is twofold. The first reason is that the particular configuration (i.e., $n_{\rm LOS}=30$ and $n_*=15$) does not have enough constraining power to perfectly identity data as coming from the correct halo. Increasing the number of stars and number of LOS would cause the maximum to increase closer to 1. The second reason is that our model uses a Dirichlet prior on the halo fractions with concentration parameters all set to 1. This means that each halo has equal prior fractions of $1/n_{\rm halo}$, and it also means that the plane implied by the halo fraction vector is flat. A consequence of this choice is that the posterior halo fractions are much more likely to propose mixtures of the halos rather than isolate a single halo as the origin, even for large numbers of LOS and stars per LOS. Since real data will never perfectly match a simulated halo, we don't expect or even want our analysis to provide only the best match to a single simulated halo.  As we show in the next section, the tendency to propose mixtures of the simulated halos gives us additional insight and constraints on halo accretion history. 

Given that our mock survey data are drawn from one particular simulated halo, we are interested in the posterior halo fraction of the correct origin halo because that tells us how distinct the data are to that halo given our observables. When it comes to future analyses of real data, the posterior halo fractions will tell us how similar the real data are to the observable distributions in each simulated halo. We know the accretion history parameters of each simulated halo, so by identifying the simulated halos that best match the observables, we can find which accretion history parameters best describe our Galaxy's halo.

We are also interested in seeing which of the other simulated halos (i.e., halos that didn't produce the data) report high posterior fractions for that data.  That implies those halos have similar observable distributions. In the next section, we show that those halos also have similarities in their accretion histories.

Our probability model effectively assumes ``the data \textit{must} originate from at least one of the halos'', but that will not be true for real data. We argue that it is still worthwhile to know which halos a set of data are most similar to, even in the case where that halo isn't a perfect representation of the data. We explore the information about halo accretion history available in these measures of similarity in S\S \ref{sec:simdist} and \S \ref{sec:charsens} below.  

While the steps above describe how we analyze data drawn from the simulated halos, the process is very similar to how we plan to analyze real data. The main difference is that we will need to account for the uncertainties on the stellar observables.  We estimate the impact of the expected \hd\ uncertainties on our results in \S \ref{sec:obserrors}. 

\section{Results and Analysis}\label{sec:results}

\subsection{Survey Design}\label{sec:num}
\begin{figure*}[ht!]
\plotone{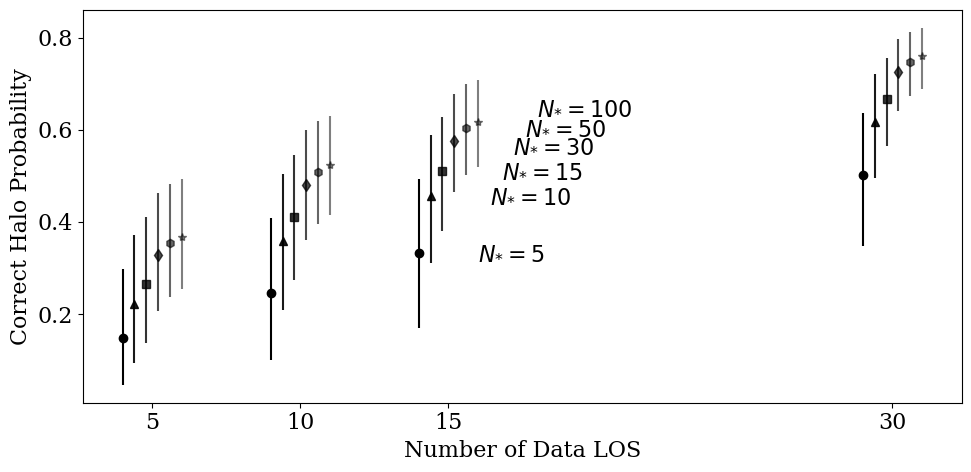}
\caption{Effect of number of data LOS and number of stars per LOS ($N_*$) on the posterior probability assigned to the halo that data were generated from. The points in each group are staggered horizontally for the sake of clarity of viewing, but they all correspond to the same number of data LOS shown on the x-axis. The y-axis is the average posterior halo probability of the halo that data were drawn from, consisting of results from all 11 halos and 360 data realizations per halo. While the correct halo probability has a slight dependence on the particular halo being considered, the variations between the halos are much smaller than the size of the errorbars shown in this figure.  A higher posterior probability indicates a better ability to distinguish the correct halo. Increasing both the number of data LOS and number of stars per LOS increase the ability to identify the correct halo. \label{fig:nlosnstars}}
\end{figure*}

\begin{figure*}[ht!]
\plotone{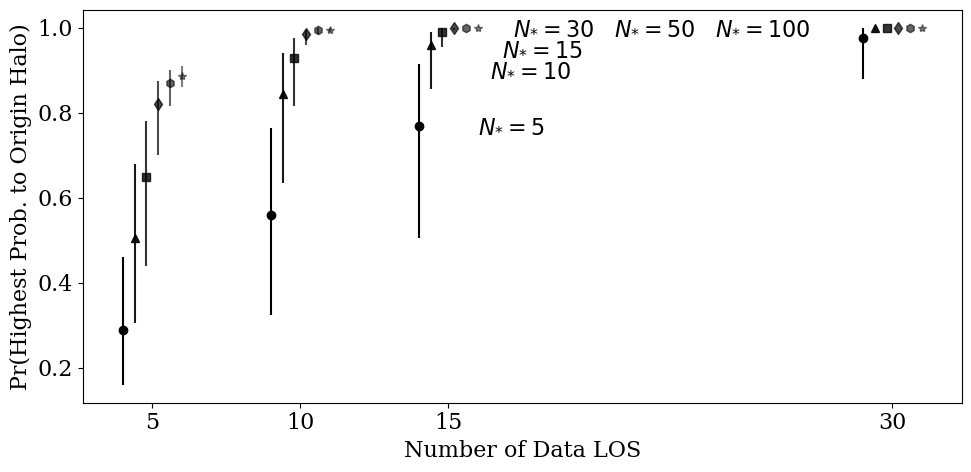}
\caption{Same as Figure~\ref{fig:nlosnstars}, but now the y-axis shows the probability that the maximum fraction in the halo probability vector is assigned to the halo that data were generated from. \label{fig:nlosnstars_maxprob}}
\end{figure*}

We repeat the analysis described in \S \ref{sec:sampeval} for all halos and for a range of survey design parameters $(n_{\rm LOS}, n_*)$ and for our baseline set of observables \vtd, distance modulus, \feh\, and \alp.  In this way, we measure the impact that our survey design choices have on the ability to match mock survey data to its parent halo. We looked at mock surveys with different numbers of data LOS (5, 10, 15, 30) and number of stars per LOS (5, 10, 15, 30, 50, 100).  We explore the results with different combinations of our observables later in \S \ref{sec:obserrors}.

Figure~\ref{fig:nlosnstars} shows the average posterior probability of the correct halo that a mock survey was drawn from for different survey design configurations. As expected, increasing the number of stars per LOS and number of LOS increases the posterior probability of the correct halo.  Figure~\ref{fig:nlosnstars_maxprob} shows the probability that the highest posterior halo fraction for a mock survey is assigned to the correct simulated halo (i.e. the probability that the values in the histograms of Figure~\ref{fig:single_halo_prob_outputs} are the highest for the correct halo). Here, we see that our technique is able to attribute the highest posterior halo fraction to the correct halo for even for small survey designs.  The results in Figures~\ref{fig:nlosnstars} and \ref{fig:nlosnstars_maxprob} show that our technique is able to identify the correct halo even with very few stars and LOS, but this is when considering an idealized test case: mock survey data drawn from one of the simulated halos. Our real \hd\ data will not match the simulated halos as closely.  In the next section, we show that, even for this idealized test case, evaluating these match probabilities for the simulated halos reveals information about the physical parameters that describe a halo's accretion history. These good results for our mock surveys give us confidence that we can identify simulated halos that have similar but not identical accretion histories to our Galaxy.

For our final \hd\ survey parameters we chose 30 LOS with at least 15 stars each, as described in \S \ref{sec:halo7d}. This maximizes our use of the available data in the HST archive, has a favorable field layout for the Keck DEIMOS spectroscopy, and includes enough stars brighter than the \Gaia\ magnitude limit along each LOS to give a high probability of matching the data to the halos with the most similar accretion histories. 

\subsection{Halo Similarities and Distinctness}\label{sec:simdist}
\begin{figure*}[ht!]
\plotone{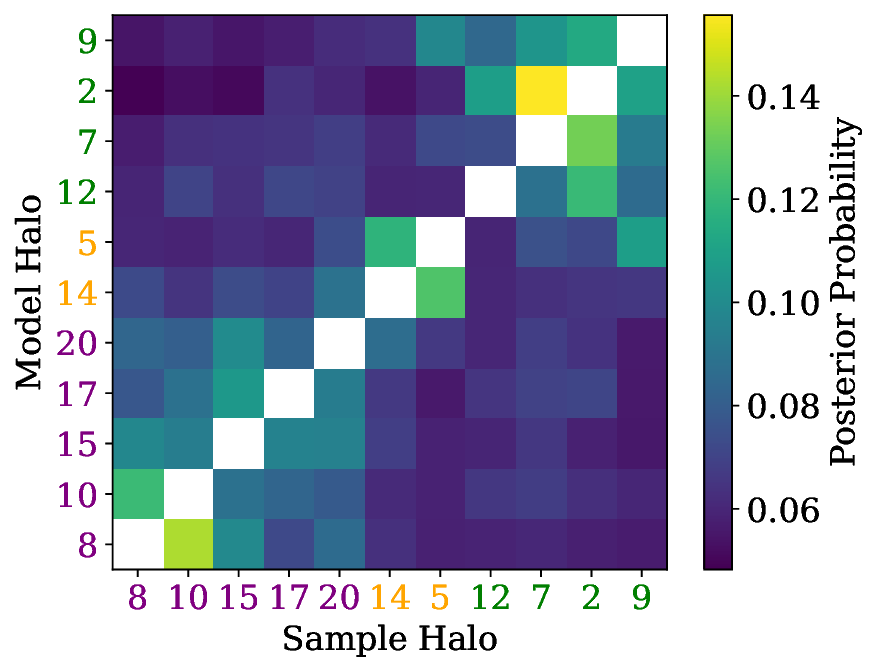}
\caption{ Summary of the evaluation of samples taken from each of the 11 halos when evaluated against all statistical models, for our \hd\ configuration of 30 LOS with 15 stars per LOS. The x-axis shows the halo the sample was drawn from, while the y-axis indicates the statistical model halo against which that sample is evaluated. Color indicates the posterior probability that the halo sample was drawn from each halo model. Values along the diagonal, where the sample was in truth drawn from that model, have been removed, as they are always the highest probability, and the remaining values have been re-scaled to sum to 1 in each column. To connect to the earlier discussion, the Halo 2  column (for example) is generated using the median posterior halo fractions (i.e., red points from Figure~\ref{fig:sampeval}), where the fractions have been re-scaled to sum to 1 after removing the Halo 2 option. The halos have been sorted to minimize the distance of high probabilities from the diagonal. Halos which had a high posterior probability of being drawn from each other were sorted together into three groups: Groups A, B, and C, as indicated by axis label color of purple, green, and gold, respectively, in order to examine the accretion history parameters that cause them to have a statistical resemblance. \label{fig:2dhist}}
\end{figure*}

In \S \ref{sec:sampeval}, we described how mock surveys drawn from one simulated halo are evaluated using the likelihood distributions of the observables for all the halos.  Figure~\ref{fig:single_halo_prob_outputs} shows the resulting posterior halo probabilities that a mock survey drawn from Halo02 -- using our baseline observables of \vtd, distance modulus, \feh\ and \alp\, and our final \hd\ configuration of 30 LOS and 15 stars each -- was drawn from each of the 11 halos.  We repeated this analysis for a set of 360 mock surveys drawn from the different Solar placement realization of Halo02.  The results are summarized in Figure~\ref{fig:sampeval}.  Halo02 has the highest posterior probability in almost all cases, meaning that our method has enough power to distinguish which halo a mock survey came from. 

However, our eventual goal is to evaluate the real \hd\ data from the MW against the simulated halos, in which case there will not be a ``correct'' halo to chose from. Therefore, we are interested in knowing not just which halo a mock survey \textit{did} come from, but which other halos it \textit{could} have come from.  We want to understand what the accretion histories of those halos have in common, and therefore what accretion history parameters we learn about by comparing \hd\ data in the MW outer halo with the \bj\ simulations.

Figure~\ref{fig:2dhist} shows the results of evaluating the set of 360 mock surveys, each with 30 LOS and 15 stars per LOS, drawn from each of the 11 halos against the probability distributions for all the halos. On the x-axis is the halo from which the mock survey is drawn, evaluated against each of the halos along the y-axis. The value in each location in the 2D array is the posterior probability that a mock survey was taken from one halo. The diagonal, which represents a match between a mock survey and the true halo from which it was drawn -- which is always the highest value -- has been removed to highlight the remaining probabilities. 

To find the most optimal sorting order of the halos along the x-axis, we score every possible way to sort the halos. For a particular sorting order, the score is calculated as
\begin{equation}
\label{eq:halo_sorting_score}
    \begin{split}
        \sum_{i} d_i^2\cdot \frac{p_i}{\sigma_{p,i}}
    \end{split}
\end{equation}
where $i$ is the index of a particular cell in the 2D histogram, $d_i$ is the minimum perpendicular distance from the center of that cell to the $y=x$ diagonal, $p_i$ is the probability value in that cell, and $\sigma_{p,i}$ is the uncertainty in that probability value. The best sorting order we present in Figure~\ref{fig:2dhist} is the ordering that had the minimum score of all the possible sorting options, which serves to move high probability values towards the diagonal and low values away from it while considering the uncertainty in those probabilities. 

From Figure~\ref{fig:2dhist}, we can see which halos resemble one another statistically in our observables, as samples drawn from one halo will have high probabilities of coming from similar halos. We put the halos in three groups, based on their probabilities of being assigned to one another: Group A with Halos 8, 10, 15, 17 and 20; Group B with Halos 12, 7, 2, and 9; and Group C with Halos 5 and 14. 

As a robustness check on our sorting, we compared the top five sorting orders as defined by their scores as well as different choices for the definition of Equation~\ref{eq:halo_sorting_score}: diagonal distance to the first power, probability to the second power, and all three of the options without the weighting by $\sigma_{p,i}$. The top five sorting options produced by the different scoring definitions all agree very closely with each other.  For every option we tested to score the halos, the best sorting order put the halos into the same three groups.  In cases where the top five sorting orders differ between the various scoring options, the difference was almost always a single pair of halos swapping positions within the same group. This suggests that our grouping of the halos by their probability scores is quite robust. 

Next, we investigate which parameters of halo accretion history create the similarities in our observables among the halos within each group, as well as the differences that make these three groups of halos distinct.

\begin{figure*}[ht!]
\plotone{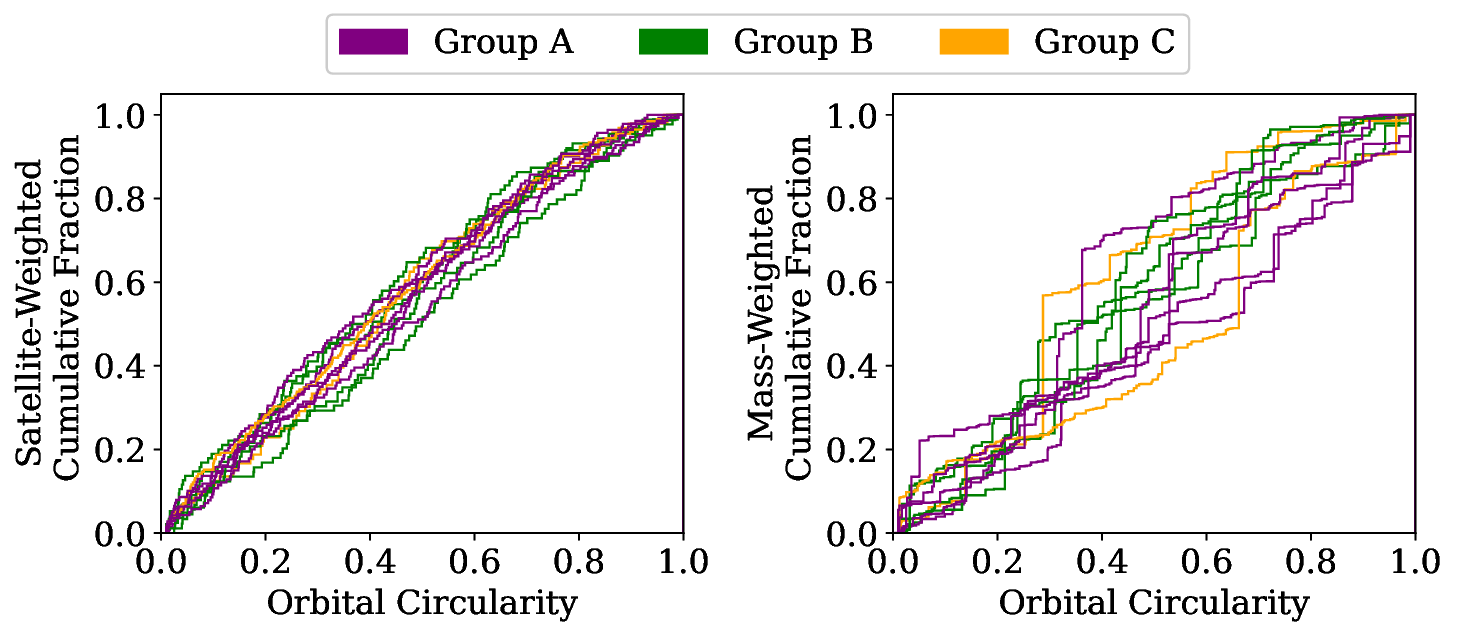}
\caption{Cumulative histograms of the orbital circularity of the satellites making up the 11 simulated halos. The left panel is unweighted, while the right panel is weighted by satellite mass. Halos are colored by groups as in Figure~\ref{fig:2dhist} and described in the text. There is no significant separation of groups by orbital  circularity, no matter the weighing. Orbital circularity of the progenitor satellite is most closely tied to the resulting morphology of the substructure created. The most likely reason for our lack of sensitivity to circularity is that our fields are too small to determine the morphology of any substructure. \label{fig:jhist}}
\end{figure*}

\begin{figure*}[ht!]
\plotone{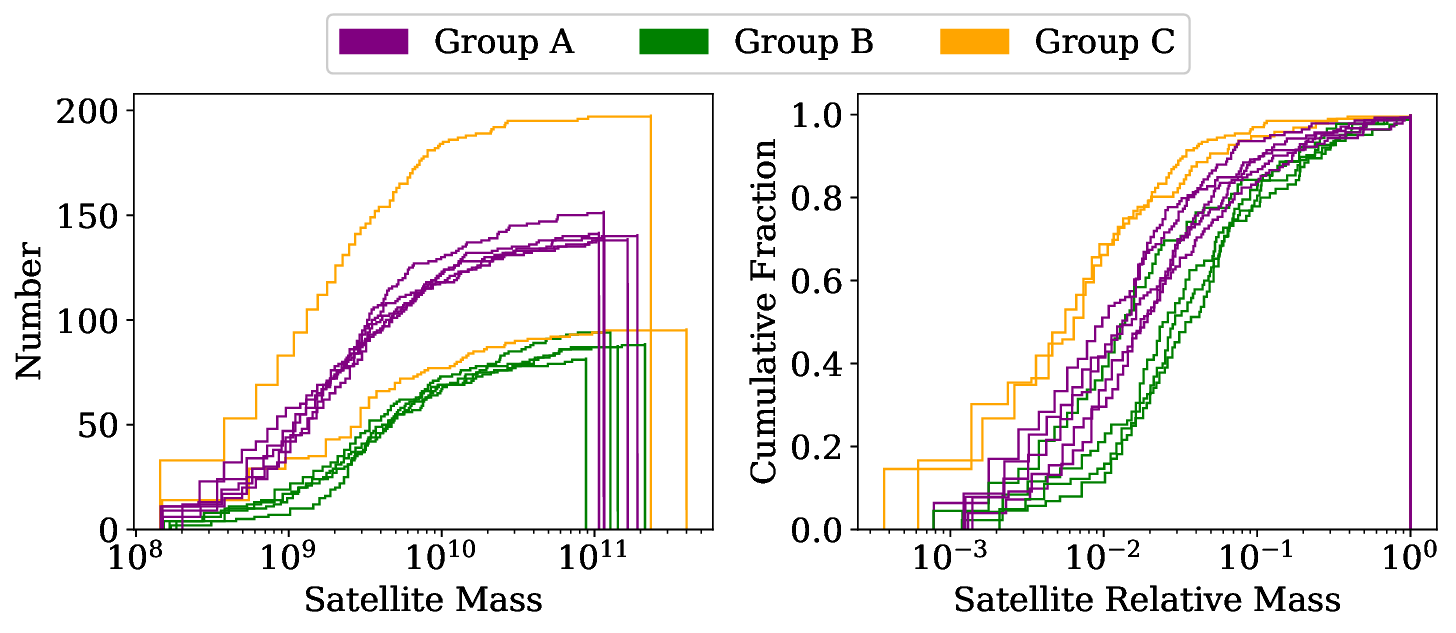}
\caption{Cumulative histograms showing the mass distribution of satellites that make up each of the simulated halos. Halos are colored by groups described as in Figure~\ref{fig:2dhist} and described in the text. The left panel is unscaled, showing the raw numbers and masses, while the right shows the cumulative fraction versus the ratio of satellite mass to the most massive satellite in each halo. In the left panel, Groups A and B are clustered together and clearly separated from each other. Group A halos were built up of many smaller satellites, while Group B had fewer small mergers, with some larger, but not significantly dominant events. Group C halos have differing numbers of small mergers but they have the two largest mergers of any halos. The two members of Group C also have the largest differences between the first and second most massive satellites. We classify these galaxies as dominated by a single massive merger event. This massive merger event means that the halos in Group C have a significantly larger fraction of their total mass made up of relatively small mergers than other halos, giving them distinct mass-weighted cumulative distributions (right panel).  \label{fig:masshist}}
\end{figure*}

\begin{figure*}[ht!]
\plotone{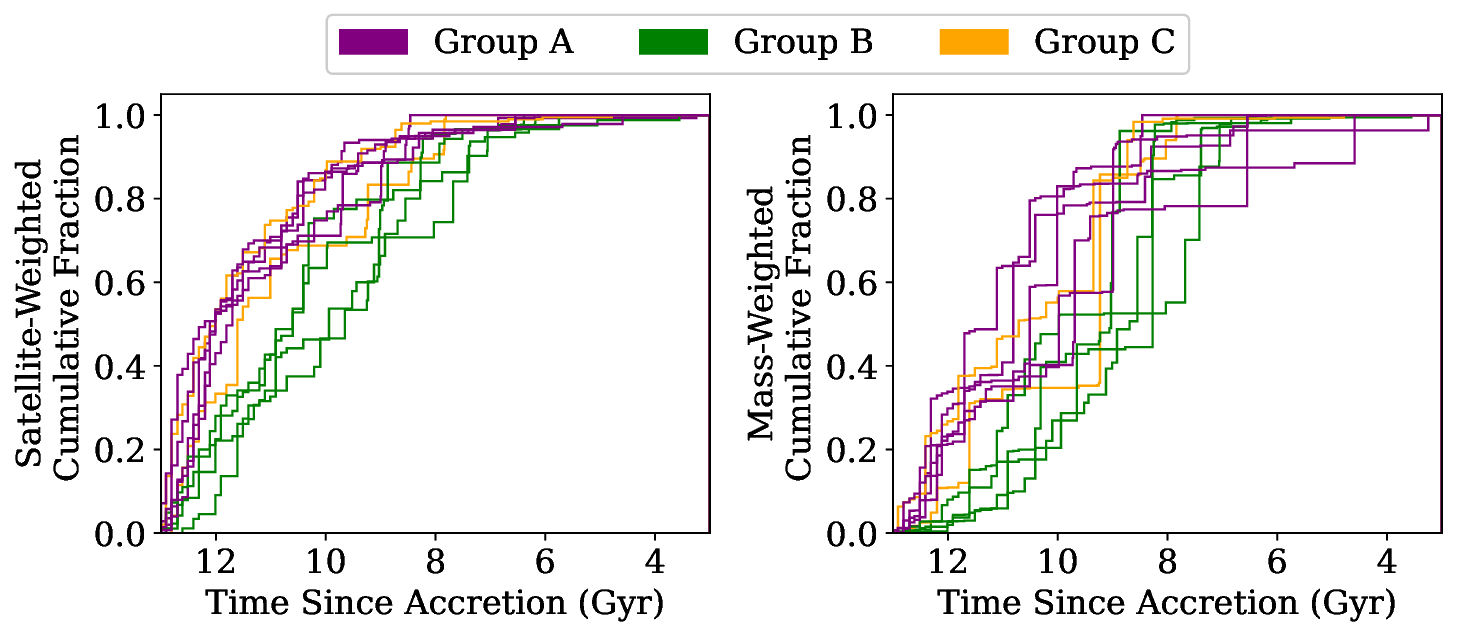}
\caption{Cumulative distributions of time since accretion for each satellite that built up our 11 simulated stellar halos. The left panel is unweighted, while the right panel is weighted by satellite mass. Halos are colored by groups described as in Figure~\ref{fig:2dhist} and described in the text. There is separation between Groups A and B throughout their accretion histories. Group A had a larger fraction of their merger events occur early in the galaxy's history, while Group B's rate of mergers has been more consistent throughout time. Group A members accreted significantly more of their mass before 10-11 Gyr ago than Group B members. Group C, like Group A, accretes many satellites early. The massive accretion events at around 9 Gyr ago distinguish the two halos of Group C visibly in these distributions, most clearly in the right panel.     \label{fig:timehist}}
\end{figure*}

\begin{figure*}[ht!]
\plotone{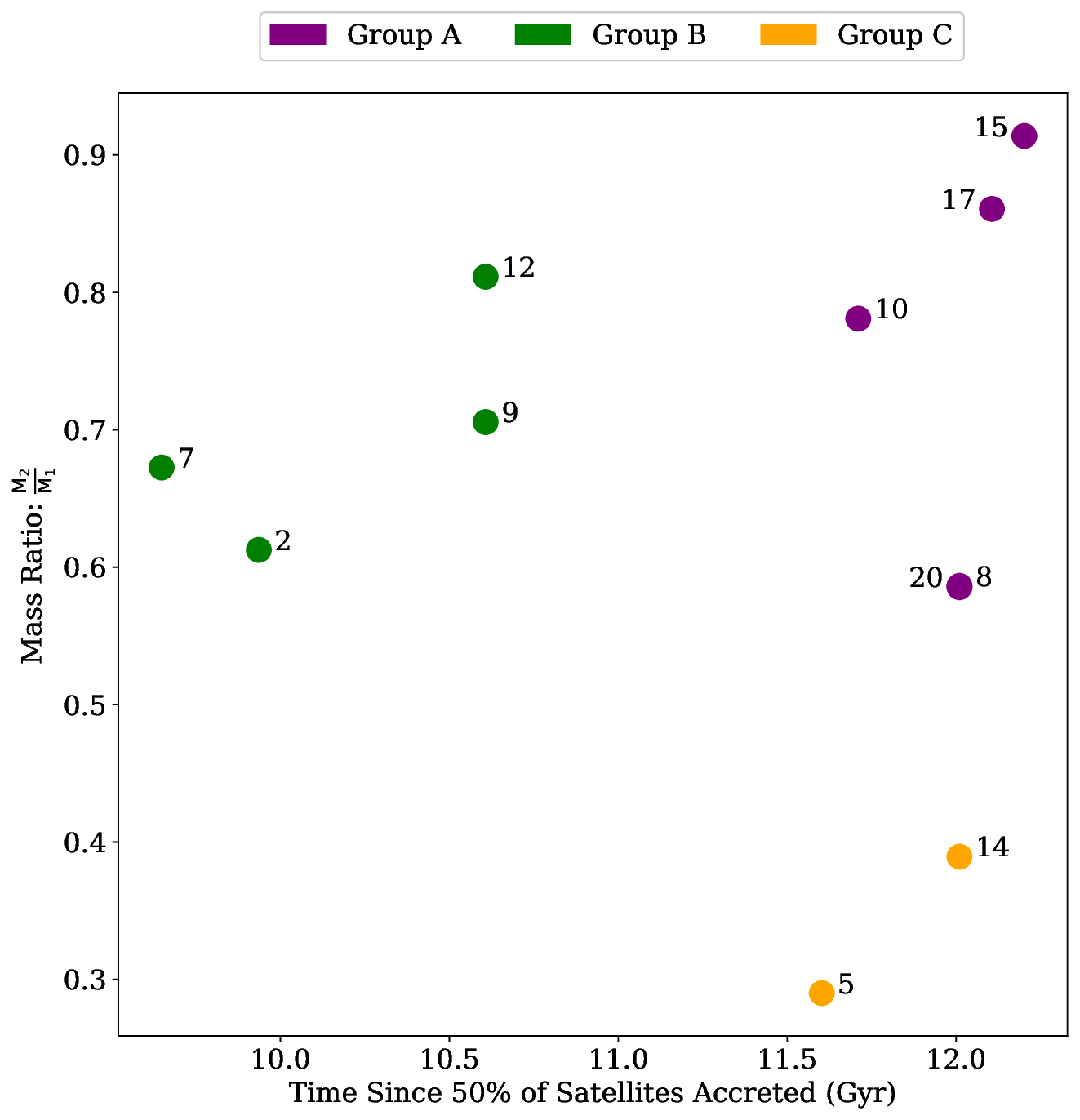}
\caption{The ratio between the masses of the two most massive satellites in each halo versus the time for 50$\%$ of satellites to accrete for the \bj\ halos. $\rm M_1$ and $\rm M_2$ are the most massive and second most massive progenitors, respectively. When looking at certain measures of mass and time together, we see all three of our groups separate from one another. Group B halos take longer for 50$\%$ of their satellites to accrete than Groups A and C. Group C is indistinguishable from Group A in time, but the ratio between the two most massive satellites is very different. In Group C halos, $\rm M_1$ is at least twice the mass of $\rm M_2$. We consider these Group C halos to have accretion histories dominated by a single major merger. Halos 8 and 20 are overlapping in this figure. \label{fig:vs2}}
\end{figure*}

\begin{figure*}[ht!]
\plotone{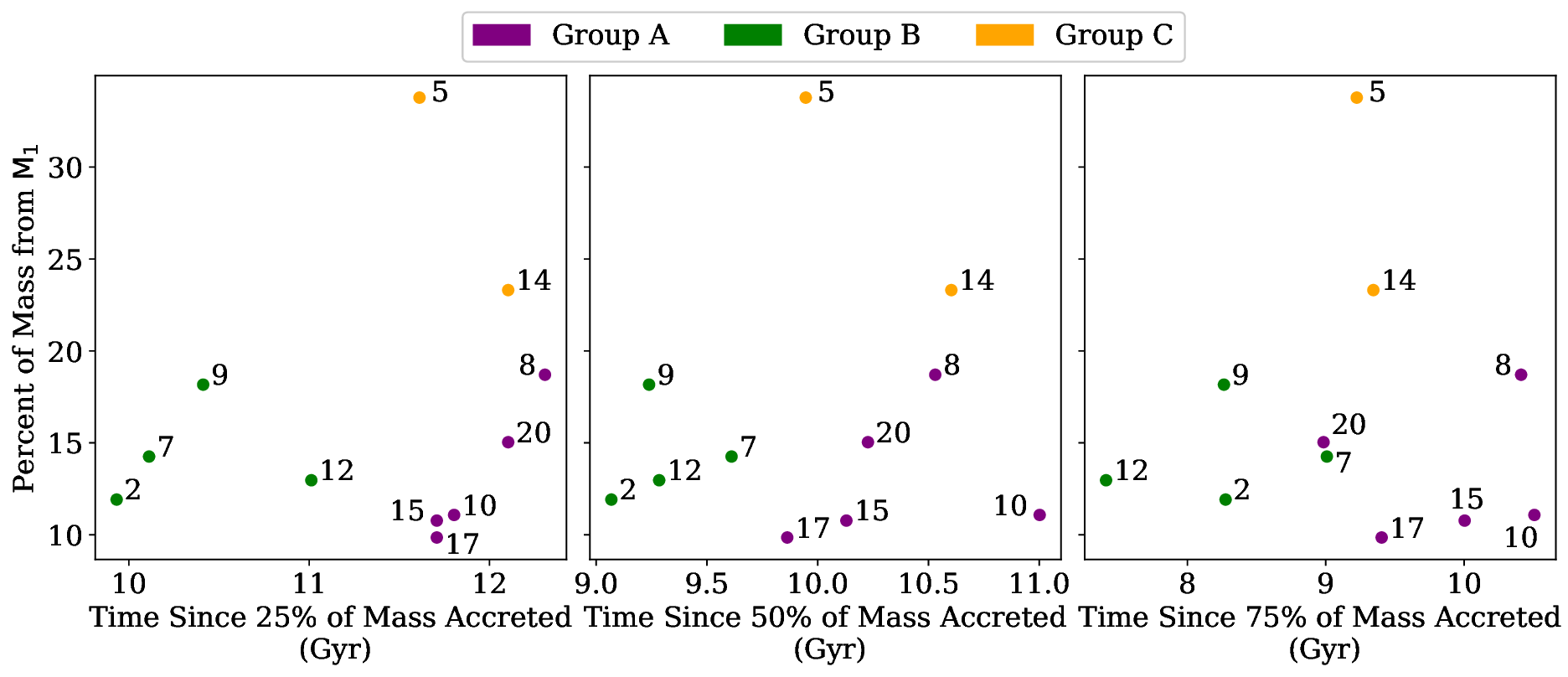}
\caption{The percent of total halo mass coming from the most massive progenitor satellite ($\rm M_1$) versus the mass accretion timeline for the \bj\ halos. Groups A and B separate out from each other at early times, with Group A halos accreting more of their mass at early times. Their distance lessens over time but the two Groups still remain distinct. Group C halos have a massive progenitor that separates them from the other halos along the y-axis. Each halo in this Group has $\rm M_1$ making up at least $20\%$ of the total halo mass.   \label{fig:vs3}}
\end{figure*}

\subsection{Sensitivity to Halo Accretion History Parameters}\label{sec:charsens}
To identify the distinguishing characteristics of the halo accretion histories for the three groups seen in the posterior halo fractions for the mock survey data shown in Figure~\ref{fig:2dhist}, we focus on the three physical parameters shown in Figure~\ref{fig:history}: orbital circularity, mass, and accretion times of the satellites that formed each of the simulated stellar halos. In Figures~\ref{fig:jhist}--\ref{fig:timehist}, we plot the cumulative distributions of these three parameters for the 11 \bj\ halos and compare them across our three groups of halos. For orbital circularity (Figure~\ref{fig:jhist}) and accretion time (Figure~\ref{fig:timehist}), the satellites are weighted by mass and number.  The satellite mass distributions  (Figure~\ref{fig:masshist}) are shown unweighted and scaled to the mass of the most massive progenitor.  In all three figures, the halos are color coded by Group as in the axis labels of Figure~\ref{fig:2dhist}. 

Looking across all three parameters, we find that the halos in Group A are formed early and most of the progenitors are low-mass satellites. The halos of Group B grow more steadily through time and are formed from fewer satellites with larger stellar masses. The two halos in Group C are dominated by a single, massive merger event, with the largest fraction of their stellar halos contributed by one progenitor.  We now consider each parameter in turn in more detail. 

For orbital circularity, shown in Figure~\ref{fig:jhist}, we see little correlation between Groups or separation between them. Our lack of sensitivity to the orbital circularity makes sense when considering our survey design. Orbital circularity is most closely tied to the resulting morphology of the substructure: satellites on more circular orbits form streams, while more radial orbits produce plumes and clouds \citep{bj08}. However, our fields are too small to contain an entire recognizable structure within them.

 The most massive contributor to the Milky Way stellar halo is the GSE progenitor satellite, which was accreted at early times, approximately 8--10 Gyr ago \citep{helmi2018}. The mass distribution of accreted satellites in the simulated halos is therefore another parameter of interest for our survey, especially the question of whether there is a single massive progenitor like GSE that dominates the accretion history.  In Figure~\ref{fig:masshist}, we show the distribution of masses of the satellites that built up each simulated stellar halo. When looking at the cumulative distribution of the number of satellites by mass in each Halo in the left panel, the halos in Groups A and B have very similar distributions within a Group, and there is clear separation between the two Groups. The halos in Group A were formed via the merging of many smaller satellites, while the stellar halos in Group B were made from the merging of fewer, more massive satellites. The two Groups of halos are also distinguishable by their total stellar mass. Although total mass, including dark matter, is similar to the MW for all the simulated halos, all Group B halos have slightly larger total masses than the Group A halos, with the largest difference being less than $10\%$. However, when comparing stellar mass, the Group B halos all have significantly larger stellar mass than the Group A halos. The average stellar mass of Group A halos is $3.2 \times 10^9$ \msun, compared to $5.6\times 10^9$ \msun\ for Group B. The progenitor satellites for Group A, on average, have lower mass than those of Group B, so they have less efficient star formation. Additionally, those low-mass progenitors were generally accreted early, which stopped their star formation entirely. These factors combine to give the Group B halos up to $75\%$ more stellar mass than Group A halos.  
 
 The two simulated halos in Group C are distinct from Groups A and B in the right panel of Figure~\ref{fig:masshist}.   The two members of Group C have the two largest progenitors by mass of any halo -- their cumulative satellite mass histograms in the left panel of Figure~\ref{fig:masshist} extend furthest to the right. The two halos in Group C also have the largest difference in stellar mass between their two most massive progenitors, and have the largest fraction of their stellar mass contributed by a single progenitor.  That makes the two halos in Group C distinct from Groups A and B in the right panel of Figure~\ref{fig:masshist}, where the cumulative mass distribution is scaled by the mass of the most massive progenitor. We describe the two halos in Group C as dominated by a single massive progenitor. 

The separation between Groups A and B is also clear when considering how stellar halo mass grows over time, as shown in Figure~\ref{fig:timehist}. Group A has a much larger fraction of the mass accreted early ($\gtrsim 11$ Gyr ago) than Group B. Group A also has many more mergers by number than Group B (left panel of Figure~\ref{fig:masshist}). The separation of the two Groups in Figure~\ref{fig:timehist} is consistent with their distinct mass functions as seen in Figure~\ref{fig:masshist}.  By the nature of hierarchical merging, halos that are built up early will naturally be made of more satellites with lower stellar mass, as those satellites will have less time to form stars before they are accreted. 

The stellar halos in Group C also have many early, low-mass accretion events and their accretion histories resemble those of the halos in Group A at early times in Figure~\ref{fig:timehist}.  However, the massive mergers at about 9 Gyrs ago that dominate the halos in Group C shift the cumulative distributions to more closely resemble those of Group B at later times (right panel, Figure~\ref{fig:timehist})

To further investigate how our grouping of the simulated halos by probabilities is related to their accretion histories, we examine the distribution of the simulated stellar halos in combinations of the parameters that summarize their accretion history. We considered summary variables such as time to 25$\%$, 50$\%$, and 75$\%$ accretion by mass and by number of satellites, and the standard deviation of the masses and the percent of mass from the largest progenitor. We looked at the distribution of the simulated halos in these summary parameters to see if we identify the same groupings found in Figure~\ref{fig:2dhist}. As expected from Figure~\ref{fig:jhist}, we found little correlation between orbital circularity parameters and our grouping of the stellar halos. Figure~\ref{fig:vs2} shows the halos in the two-dimensional plane of satellite mass distribution and accretion time. Here we see the halos separate into the same groups as we found in Figure~\ref{fig:2dhist}.  The Group A halos accrete 50\% of their mass before 11.5 Gyr ago, while the Group B halos reach 50\% of their stellar mass between 10 and 10.5 Gyr ago.  Group C separates out along the y-axis, which shows the ratio between the two most massive satellites, a measure of the dominance of a single merger event. We show the halos in a related set of time and mass fraction variables in Figure~\ref{fig:vs3}.  Here the horizontal axes show the distribution of the halos at three different points in their cumulative mass accretion. The halos in Group A accrete 25\%, 50\%, and 75\% of their stellar mass at later times than the halos in Group B. This separation between the Groups is more distinct at earlier times. The Group C halos stand out from Group A in these figures on the y-axes, which plot the fractional contribution from the largest satellite, as a measure of mass distribution of the satellites. 

We assigned the simulated \bj\ stellar halos as observed with our HALO7D-X survey design into Groups A, B, and C based on their similarity in the probabilistic comparisons of their observables described in \S~\ref{sec:statmodels} and shown in Figure~\ref{fig:2dhist}. The halos also separate into the same groups when we compare their known accretion histories in the two-dimensional plane of parameters that describe the accretion time and mass distribution of their progenitors, as shown in Figures~\ref{fig:vs2} and \ref{fig:vs3}. The halos in the three groups also follow similar cumulative measurements of mass and accretion time in Figures~\ref{fig:timehist} and \ref{fig:masshist}. This gives us confidence that our survey design is sensitive to the mass distribution and accretion timeline of the satellite progenitors of our Galaxy's stellar halo. It also demonstrates that we can use our probabilistic comparison between data from our HALO7D-X survey and the set of simulated halos as a measure of the similarity of the stellar halo's accretion history to its closest matches among the set of simulated halos. 

\subsection{Impact of Observational Uncertainties and Observables}\label{sec:obserrors}

\begin{figure*}[ht]
\plotone{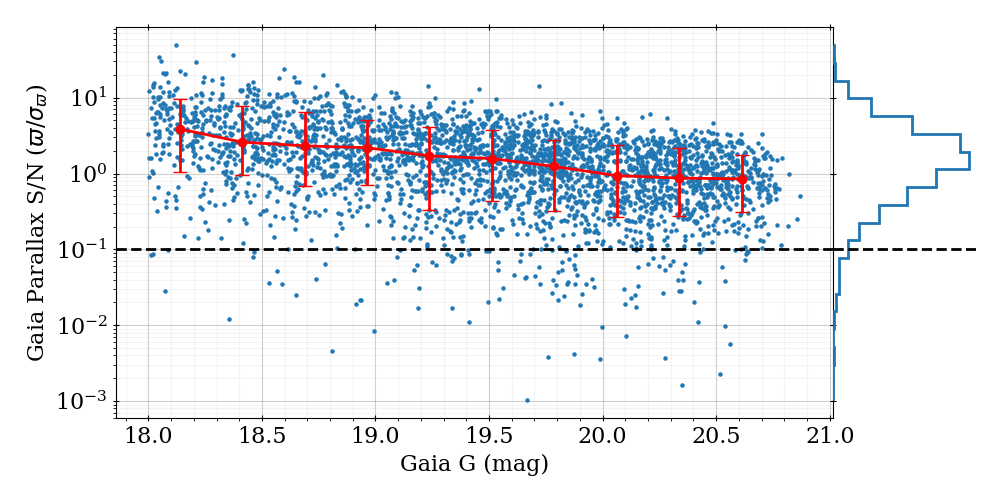}
\caption{Gaia parallax S/N as a function of magnitude for the $3128/4213\approx74\%$ of COSMOS stars shown in Figure~\ref{fig:COSMOS_CMD} that have Gaia-measured parallaxes (blue points). Medians and 68\% intervals are shown in red for different magnitude bins, and a histogram of the parallax S/N is shown on the right edge. 96\% of the blue points lie above S/N$=0.1$, which is the limit for which we expect to measure posterior spectrophotometric distances with uncertainties smaller than $\sim35\%$. \label{fig:COSMOS_parallax_SNR}}
\end{figure*}

So far we have demonstrated that our measures of similarity between our mock survey observations and the set of simulated halos reflect real similarities in the halo accretion histories. The distributions of the observables we used for the comparisons were constructed with the grid resolution and smoothing scales shown in Table~\ref{tab:observables_grid}. As stated in \S \ref{sssec:pdfs}, the smoothing scale for constructing those distributions was chosen to match the distribution of the observables over the set of simulated halos. We can now examine the expected impact of our estimated observational uncertainties and choice of observables on our ability to identify halos with similar accretion histories.  

To estimate the observational uncertainties for our \hd\ observations, we use the original HALO7D data scaled to our \hd\ survey design as appropriate. Our \hd\ observations will have a brighter limiting magnitude of $G= 21.5$~mag than the original HALO7D data, but we will use correspondingly shorter integration times.  

For the velocity probability distributions, \vtd\ and \vlos, we used a grid cell size of 200 km~s$^{-1}$ and 120 km~s$^{-1}$ respectively, and for both velocity observables we used a smoothing scale of 60 km~s$^{-1}$. The cell sizes and smoothing scale are significantly larger than our expected velocity uncertainties.  Based on the HALO7D velocities described in \citet{c19a} we expect the line of sight velocity uncertainty at our magnitude limit of 21.5 to be 10 km~s$^{-1}$.  Our \vtd\ measurements also include the proper motions.  The deep HST+\Gaia\ proper motions we will use for \hd\ are described in \citet{mckinnon23b}. That analysis predicts a proper motion uncertainty of 1.16 mas yr$^{-1}$ per coordinate at $G$ of 21.5.  The median heliocentric distance of the HALO7D chemistry sample is 8 kpc \citep{mckinnon23a}.  At that distance the HST+\Gaia\ proper motion errors at $G$ = 21.5 correspond to a transverse velocity error of 44 km~s$^{-1}$.  The resulting 3D velocity uncertainty is 63 km~s$^{-1}$, very close to the smoothing scale and much smaller than the grid size for our probability distributions, so not likely to change the results presented here.  Because the 3D velocities include our distance estimates, we will need to include the correlated uncertainties in distance and 3D velocity in our analysis of the real \hd\ data. To estimate the impact of distance errors we repeated our analysis for Figure~\ref{fig:2dhist} using LOS velocity (\vlos) instead of \vtd. This represents an extreme dilution of the information available in the transverse velocities in the case of very large distance errors.   We found that the halos sorted into the same groups we found in Figure~\ref{fig:2dhist} using \vtd, suggesting that we will not see a major change to our results when we include these covariances. 

The cell size for the distance modulus is 2 magnitudes and the smoothing scale is 0.75 magnitudes. This smoothing scale corresponds to a fractional distance uncertainty of $\sigma_{D}/D = \sigma_{DM}\cdot\ln(10)/5 ~= 35\%$. The grid size of 2 magnitudes is equivalent to an 92\% distance uncertainty. As a result, we do not expect a significant change to our results when including the distance errors. Even though the errors are large, the relative distances remain correct on average, retaining information about the radial distribution of halo progenitors.

For the spectrophotometric analysis of real \hd\ data, we will likely recover much more precise distances than we've assumed here. This is because we will have access to prior distance information from astrometric parallaxes (either from Gaia alone or HST+Gaia measurements), which was not available for the analysis of much fainter stars in the original HALO7D catalog \citep{mckinnon23a}. \hd\ will estimate distances from a combination of spectra, photometry, and Gaia-informed parallaxes. This means that posterior distance uncertainties will more closely match the outputs of similar pipeline like \texttt{MINESweeper} \citep{Cargile_2020}. From Figure 3 of \citet{Cargile_2020}, fractional uncertainties on posterior distances are almost always much smaller than 35\% for a range of evolution phases (main sequence, turnoff, and red giant branch) and \feh, even for low S/N spectra (S/N$_{\mathrm{spec}} = 2$) and extremely low S/N parallax priors ($\varpi/\sigma_{\varpi} = 0.1$). Looking at the same example COSMOS sample from Figure~\ref{fig:COSMOS_CMD}, the vast majority of potential \hd\ stars with parallax measurements surpass this low S/N threshold, as demonstrated in Figure~\ref{fig:COSMOS_parallax_SNR}.

For \feh, the grid size is 0.2 dex and the smoothing scale 0.05 dex.  We expect a median S/N ratio of about 20 for our \hd\ observations.  Based on the HALO7D chemistry results of \citet{mckinnon23a} the corresponding median uncertainty in \feh\ is 0.22 dex, about equal to our cell size.  We will evaluate the impact of the true \hd\ uncertainties on our analysis of the real data once we know them (see below on pipeline improvements). But this scaling suggests that the  \hd\ \feh\ uncertainties are likely to contribute some additional smoothing to the probability distributions, but not change them significantly.  

 Finally, we consider \alp. At our estimated median S/N ratio for \hd\ the \alp\ uncertainties from the HALO7D chemistry analysis of \citet{mckinnon23a} are 0.4 dex, significantly larger than the grid size we used in this analysis for \alp\ of 0.09 dex and our smoothing scale of 0.025 dex. Our grid parameters and smoothing scale were chosen to reflect the information content in the simulations.  To get a sense of the impact of the \alp\ observational uncertainties, we repeat our analysis in the extreme case of removing \alp\ entirely and using only the remaining three observables: \vtd, distance modulus, and \feh.  We find that the halos sort into the same Groups A and B, with the same membership. But in this case Halos 5 and 14 are no longer distinct as Group C.  Halo 5 sorts into Group B, the Group with halos made from fewer mergers with larger satellites. This is as expected, since the halos in Group C are dominated by one massive merger.  Halo 14 sorts into Group A, which implies that even though like Halo 5 it has a very massive progenitor, the fact that it also has many early accretion events (see Figure~\ref{fig:history}) has a stronger impact on the observables.  That makes Halo 14 a better match to Group A, which is comprised of halos with predominantly early accretion, than to the halos of Group B with more massive progenitors accreted later.  
 
 In summary, when we analyze data with real uncertainties in the observables we don't expect to lose much information in the velocity and distance modulus dimension, and we expect only moderate additional smoothing for \feh.  For \alp\ we may lose sensitivity to some aspects of halo accretion history. From the estimates presented here that will most likely be less leverage on the mass distribution of halo progenitors.  However, we are optimistic that we can improve our chemistry analysis for \hd\ and reduce the \feh\ and \alp\ uncertainties. Unlike the original HALO7D, our entire \hd\ sample will have \Gaia\ parallax measurements, improved by combining with HST as described in \citet{mckinnon23b}. This will give us significantly stronger constraints on priors for $\log{g}$ and distance, which should in turn improve the measurements of \feh\ and \alp, as demonstrated by \citet{Cargile_2020}.  We also plan to improve on \citet{mckinnon23a} by using isochrones that span a range of \alp.  These improvements should reduce the \feh\ and \alp\ uncertainties and recover some of the information in our chemistry observables.

\section{Conclusions}\label{sec:end}
We present the design for the \hd\ survey of the stellar halo of our Milky Way Galaxy. \hd\ will use positions and proper motions of faint stars ($G<21.5$~mag) measured by combining  HST and \Gaia, and Keck spectroscopy to measure line-of-sight velocity, distance, \feh, and \alp. Combined, \hd\ will have three-dimensional positions, velocities, and chemistry for halo stars to study the accretion history of our Galaxy's stellar halo.  The final design for \hd, described in \S \ref{sec:num}, covers 30 LOS, each consisting of multiple archival HST fields and optimized for Keck DEIMOS spectroscopy.  We expect to find at least 15 halo stars in each LOS. 

We use the \bj\ simulated stellar halos to determine the sensitivity of our survey to the main parameters of the MW's accretion history. We use the 11 simulated halos to create probability mass functions for each of our observables, and combine them to build multidimensional probability distributions for our observables. We draw mock HALO7D-X surveys from the simulated halos and compare them to the probability distributions for the set of simulated halos. We find  that the mock surveys are reliably matched to their original halos with the highest posterior probability. In addition, the mock surveys have significant posterior probability of matching to simulated halos other than the one they are drawn from, indicating that those halos are similar terms in our observables (\S \ref{sec:simdist}). 

We show that the set of halos that are high probability matches to a mock survey in the observables also share similarities in the parameters that describe their accretion histories (\S \ref{sec:charsens}). In our mock \hd\ analysis, the simulated stellar halos fall into one of three groups, and the members of each group have common features in their accretion histories: 
\begin{enumerate}
\item[Group A:]
Stellar halos built up early of many small mergers (mean mass $\approx 7.5 \times 10^9$  \msun).
\item[Group B:]
Stellar halos built up steadily over time, with a smaller number of larger, but not dominant, mergers (mean mass $\approx 11 \times 10^9$  \msun).
\item[Group C:]
Stellar halos with a major merger (satellite mass $\geq$ 20\% of total mass) that dominates the accretion history and makes them observably distinct from Groups A and B. 
\end{enumerate}

We conclude that the \hd\ survey will be sensitive to both the mass distribution and the timeline of accretion of the Milky Way's stellar halo progenitor satellites. However, \hd\ will not be able to determine the orbital properties of the progenitors.

Ultimately, \hd\ will compare real Milky Way data with the simulated halos, using the probability distributions of our observables as described in \ref{sec:sampeval}.  For the real \hd\ data, there will be no ``true'' simulated halo to match. But we have shown that the posterior probabilities that result from comparing the \hd\ observables to the simulated halos can be used to identify which of those simulated halos the survey \textit{could} have been drawn from, and what about their accretion histories those halos have in common with one another and with the Milky Way. From our analysis of the mock surveys drawn from the \bj\ halos, we expect \hd\ to be most sensitive to measurements of the mass distribution of the progenitors and the timeline of their accretion.

\begin{acknowledgments}
Funding was provided by NASA through grants associated with HST Archival Program AR-16625 awarded by the Space Telescope Science Institute, which is operated by the Association of Universities for Research in Astronomy, Inc., for NASA, under contract NAS5-26555. KM was also supported by the Natural Sciences and Engineering Research Council of Canada (NSERC), [funding reference \#DIS-2022-568580]. 

\end{acknowledgments}

\software{Astropy \citep{astropy13, astropy18, astropy22}, Galaxia \citep{galaxia}}

\bibliography{main}{}
\bibliographystyle{aasjournal}

\end{document}